%% file: main.tex
\documentclass[pdflatex,sn-mathphys-num]{sn-jnl}
\usepackage{graphicx}%
\usepackage{multirow}%
\usepackage{amsmath,amssymb,amsfonts}%
\usepackage{amsthm}%
\usepackage{mathrsfs}%
\usepackage[title]{appendix}%
\usepackage[dvipsnames]{xcolor}%
\usepackage{textcomp}%
\usepackage{manyfoot}%
\usepackage{booktabs}%
\usepackage{algorithm}%
\usepackage{algorithmicx}%
\usepackage{algpseudocode}%
\usepackage{listings}%
\usepackage[utf8]{inputenc}

\usepackage{textgreek}
\usepackage{cleveref}
\usepackage{listings}

\definecolor{primary}{RGB}{0, 51, 102}    
\definecolor{secondary}{RGB}{230, 240, 255} 
\definecolor{accent}{RGB}{100, 100, 100}  

\definecolor{keywordcolor}{rgb}{0.7, 0.1, 0.1}   
\definecolor{tacticcolor}{rgb}{0.0, 0.1, 0.6}    
\definecolor{commentcolor}{rgb}{0.4, 0.4, 0.4}   
\definecolor{symbolcolor}{rgb}{0.0, 0.1, 0.6}    
\definecolor{sortcolor}{rgb}{0.1, 0.5, 0.1}      
\definecolor{attributecolor}{rgb}{0.7, 0.1, 0.1} 

\theoremstyle{thmstyleone}%
\newtheorem{theorem}{Theorem}
\theoremstyle{thmstyletwo}%

\theoremstyle{thmstylethree}%
\newtheorem{definition}{Definition}%

\usepackage{xspace}
\newcommand{\cslib}{CSLib\xspace}

\newcommand{\lstlabel}[1]{\hypertarget{lst:#1}{}}
\newcommand{\lstref}[1]{\hyperlink{lst:#1}{\lstinline|#1|}}

\usepackage{pgfplotstable}

\usepackage{booktabs}

\usepackage{cleveref}
\usepackage{todonotes}
\usepackage{etoolbox}

\makeatletter

\def\lst@endchar{}
\lst@Key{end}{}{\def\lst@endchar{#1}}

\newcommand{\lst@putendchar}{%
  \ifx\lst@endchar\@empty
  \else
    \nobreak\hfill\mbox{\normalfont\lst@endchar}%
  \fi
}

\patchcmd{\lst@DeInit}
  {\lst@XPrintToken \lst@EOLUpdate}
  {\lst@XPrintToken \lst@putendchar \lst@EOLUpdate}
  {}
  {\PackageError{listings-end}{Could not patch \string\lst@DeInit}{}}

\usepackage{newunicodechar}
\newunicodechar{⇓}{$\Downarrow$}
\newunicodechar{⊨}{$\vDash$}
\newunicodechar{⊤}{$\top$}
\newunicodechar{ᶜ}{${}^{\texttt c}$}
\newunicodechar{◇}{$\Diamond$}
\newunicodechar{□}{$\Box$}
\newunicodechar{⊓}{$\sqcap$}

\makeatother

\input{macros}

\begin{document}

\title[Library-Grade Modal Logic]{Library-Grade Modal Logic}

\author[1,2]{\fnm{Marianna} \sur{Girlando}}

\author[1]{\fnm{Fabrizio} \sur{Montesi}}

\affil[1]{\orgdiv{FORM}, \orgname{University of Southern Denmark}, \orgaddress{\street{Campusvej 55}, \city{Odense}, \postcode{5230}, \state{Southern Denmark}, \country{Denmark}}}

\affil[2]{\orgdiv{ILLC}, \orgname{University of Amsterdam}, \orgaddress{\street{Science Park 103}, \city{Amsterdam}, \postcode{1098XG}, \country{Netherlands}}}


\abstract{%
Modal logic comprises a broad family of logics to reason about relational structures. 
The literature presents many such logics, displaying diverse operators, semantics, and applications.
This plurality is reflected in a zoo of mechanised modal logics, which often duplicate syntax, semantics, metatheory, and reasoning infrastructure.

We present a library-grade formalisation of modal logic in Lean, developed as part of CSLib (the Lean Computer Science Library) and designed around two complementary forms of reuse: vertical reuse, whereby specialised logics inherit from common abstractions, and horizontal reuse, whereby modal logic becomes a reasoning tool for independently formalised domains. Our development provides a generic framework for polyadic modal languages, reusable metatheory and proof automation, and derived interfaces for specialised modal logics. We exemplify our infrastructure by deriving basic modal logic, basic temporal logic, and Hennessy–Milner Logic; the latter subsumes and extends CSLib's previous implementation while reducing its logic-specific code by 67\%. We further apply the same modal infrastructure to reasoning about mathematics (radicals of ideals), theory of programming languages (the type safety strategy for the simply typed $\lambda$-calculus), and concurrency theory (reasoning about processes in the Calculus of Communicating Systems). Our experience suggests that reuse, automation, and integration with the strong Lean ecosystem should be treated as first-class design concerns when formalising theories for shared libraries.}




\maketitle


\section{Introduction}\label{sec:intro}

\begin{quote}
`Ask three modal logicians what modal logic is, and you are likely to get at least three different answers.'
\\
-- Patrick Blackburn, Maarten de Rijke, Yde Venema, \emph{Modal Logic} \cite{BRV01}.
\end{quote}

This article is about the formalisation of modal logic. In particular, we aim at developing a reusable library that others can build on, thus accommodating the usefulness and plurality of modal logic itself.

Modal logic can be seen as a long-studied family of logics for reasoning about relational structures. 
Its distinctive feature is the addition of modal operators, which are not truth-functional, meaning that truth of a modal proposition cannot be established merely as a function of the truth values of its subpropositions. Instead, truth of a modal proposition at a world (or state) depends on the truth of propositions at other worlds related to it.
Familiar examples are \emph{possibility} -- stating that a proposition holds at some related state -- and \emph{necessity} -- stating that a proposition holds at all related states \cite{BRV01}.
Different choices of modal operators, relations, and properties of those relations give rise to a remarkably broad family of logics, with applications throughout mathematics, philosophy, and computer science.

While this plurality is a key strength that makes modal logic interesting, it also poses a challenge for formalisation.
One can formalise each modal logic separately, giving it its own syntax, semantics, metatheory, and reasoning infrastructure.
Indeed, the current landscape of mechanised modal logic consists of a variety of largely independent developments, spanning different modal systems and theorem provers \cite{BGHRDV26,BZ25,OBKS24,B22,lentil-github,FBV20,BCHBPT22,B21,DS11}.
Taken together, these developments form a zoo of mechanised modal logics, raising the question of how much of their common structure can instead be reused.

This question arose concretely in the development of \cslib \cite{cslib,cslib-spine}, the computer science library of the Lean programming language and proof assistant \cite{MU21}.
In the area of logic, \cslib aims at formalisations that, beyond being interesting for the study of the object of formalisation itself, offer reusable reasoning infrastructure that can support both the development of specialised theories and applications to independently formalised domains.
For modal logic specifically, pursuing such a \emph{library-grade} ambition means seeking reuse in two complementary directions.
First, specialised modal logics should be obtainable from common abstractions, inheriting as much syntax, semantics, metatheory, and reasoning infrastructure as possible -- a form of \emph{vertical reuse} \cite{D93}.
Second, modal logic itself should be available as a reasoning tool for independently formalised mathematical and computational structures -- a form of \emph{horizontal reuse} \cite{D93}.

In this article, we develop such a library-grade architecture for modal logic in Lean. Our main contributions are:

\begin{itemize}
\item A generic framework for polyadic modal logic, featuring a parametric modal language and semantics (\Cref{sec:framework}). To our knowledge, this is the first formalisation of the textbook approach from \cite{BRV01} to a generic, parametric modal language.
We further generalise the approach from \cite{BRV01}: instead of assigning each operator a finite arity $n$, we assign it an arbitrary type of argument positions. This subsumes ordinary $n$-ary operators, but also allows non-numerical and infinite collections of arguments. We represent such signatures as polynomial functors.

\item Reusable metatheory and automation, including novel generalised frame conditions 
and their corresponding axioms, logical equivalence, contexts and congruence, denotational and theory-based reasoning, and an extensible \lstinline|modal| database for automated reasoning through Lean's recent \lstinline|grind| tactic \cite{MM26} (\Cref{sec:metatheory}).

\item A hierarchy of typeclasses for derived modal interfaces and logics, including unary and unimodal interfaces (\Cref{sec:derived-modal-operators}).
From this hierarchy, we derive modal logics of different complexity such as basic modal logic (covering the entire modal cube), basic temporal logic, and Hennessy--Milner Logic ($\HML$), all inheriting reusable metatheory and infrastructure from the generic layer (\Cref{sec:deriving-modal-logics}).
Our implementation of $\HML$ completely subsumes and extends the previous ad hoc implementation in \cslib \cite{MPR26}, including its Hennessy--Milner theorem that identifies logical equivalence with bisimilarity, while reducing the logic-specific $\HML$ codebase by 67\% (\Cref{table:modal-reuse}).

\item Interfaces for applying modal reasoning to independently formalised Lean developments (\Cref{sec:applied-modal-reasoning}), demonstrated through case studies in mathematics (ideals and radicals), theory of programming languages (type safety for the simply typed $\lambda$-calculus), and concurrency theory (verification of processes in Milner's CCS).
Notably, our design benefits from prioritising the adoption of and integration with appropriate APIs available in the Lean ecosystem for mathematics -- from the Mathlib library \cite{mathlib} -- and computer science -- \cslib.
\end{itemize}

In \Cref{sec:discussion}, we discuss what lessons can be gained from our development about (i) modal logic itself and (ii) library-grade formalisation in general, as well as 
directions for future work. We then conclude in \Cref{sec:conclusion}.

In the sections that follow, we assume that the reader is familiar with Lean and some basic notation from Mathlib (like set-builder notation). A recap of the relevant background on modal logic is given in \Cref{sec:background}.

\paragraph{Reproducibility and Availability} All code has been verified using Lean 4.35.0-rc3. The implementation described in this paper is available at the \cslib project.\footnote{\url{https://github.com/leanprover/cslib}. See commit \href{https://github.com/leanprover/cslib/commit/0d7538fb3da6dbf2e050b4fb084068822b93c7af}{\texttt{0d7538fb3da6dbf2e050b4fb084068822b93c7af}} for the generic framework, basic modal logic, and HML; \url{https://github.com/leanprover/cslib/pull/927} for basic temporal logic; and \url{https://github.com/leanprover/cslib/pull/904} for the application to CCS.}

\paragraph{Related works}
As we have already mentioned, the landscape of formalisations of modal logic is vast, yet largely fragmented \cite{BGHRDV26,BZ25,OBKS24,B22,lentil-github,FBV20,BCHBPT22,B21,DS11}.
To our knowledge, no existing development pursues a library-grade formalisation as intended here, although several developments provide reusable machinery or expressive general modal languages \cite{OBKS24,OML26}.
Our approach is different from \cite{OBKS24,OML26} from the outset, 
since we base our development on a further generalisation of the generic modal language from \cite{BRV01} 
with polynomial functors, and build on that to obtain a hierarchy of abstractions and new general notions on frames that support the development of specialised logics without requiring metaprogramming. This is important, among other reasons, because it keeps the construction of specialised modal logics simple.
Our applications (\Cref{sec:applied-modal-reasoning}), approach to automation, and development of logical equivalence are also new.
Nevertheless, reconstructing features from formalised coalition logic \cite{OBKS24} and hybrid logic \cite{OML26}, as well as their proof systems, is an interesting future direction.

\section{Background}
\label{sec:background}
\label{sec:background:modal}
In this section, we recap the necessary background on modal logic and the textbook generic framework for modal languages \cite{BRV01}. All the definitions in this section are taken from \cite{BRV01}.

\begin{definition}
\label{def:modal:sim:type}
    A \emph{modal similarity type} is a pair $\modsim = \langle O, \rho \rangle $, where $O$ is a nonempty set of modal operators, denoted by $\tri_1, \tri_2, ..$, and $\rho : O\longrightarrow \nat$ is a function which assigns to every $\tri \in O$ a finite \emph{arity}, which indicates the number of arguments that $\tri$ takes.
\end{definition}

We call $\tri$ a `triangle', following \cite{BRV01}.
When $O$ contains only unary triangles, we call them \emph{diamonds} and write $\DIAp{i}\varphi$ for $\tri_i(\varphi)$. Furthermore, if $O$ consists of exactly one unary triangle $\tri$, we write $\DIA\varphi$ for $\tri(\varphi)$.

\begin{definition}[Generic modal language]
\label{def:mod:language}
    Given a countable set $\Prop$ of propositional variables and a modal similarity type $\modsim = \langle O, \rho \rangle $, a \emph{modal language} $\modL{\tau, \Prop}$ consists of the formulas generated as follows, where $p \in \Prop$ and $\tri \in O$: 
    $$
    \varphi ::= p \mid \bot \mid \NEG \varphi \mid \varphi \OR \varphi \mid  \tri(\varphi_1, \dots, \varphi_{\rho(\tri)})
    $$
\end{definition}

Other propositional operators are derived in the standard way, e.g., $\varphi \to \psi:= \neg \varphi \lor \psi$. Moreover, for every $\tri_i \in O$, we define its dual $\trid_i$ (`\emph{nabla}') as $\trid_i (\varphi_1,\dots ,\varphi_n):= \NEG \tri_i(\NEG \varphi_1, \dots, \NEG \varphi_n)$. A unary $\trid_i$ is called \emph{box}. 
As before, we write $\BOXp{i}\varphi$ for $\trid_i(\varphi)$, and we write simply $\BOX \varphi$ if additionally $O$ consists of only one operator.

Many well-known modal languages are special cases of the generic modal language \cite{BRV01}.
For example, the \emph{basic modal language} is defined by the modal similarity type $\modsim_0 = \langle O_0, \rho_0\rangle$, where $O_0 = \{\DIA \} $ and $\rho_0 (\DIA) = 1$.

We now define the semantics of modal languages, by specifying the notions of frames and models. These notions will allow us to evaluate modal formulas, thus establishing their \emph{satisfiability}. 

\begin{definition}[$\modsim$-frame]
\label{def:frames}
    Let $\modsim = \langle O,\rho \rangle$ be a modal similarity type. A \emph{$\modsim$-frame} is a tuple 
    $$
    \fr = \langle W, \{R_\tri \mid \tri \in O \text{ and } \rho(\tri )> 0\}\rangle 
    $$ 
    where $W $ is a nonempty set of elements called \emph{worlds}, or \emph{states}, and, for each $n$-ary operator $\tri \in O$ such that $n\geq 0$, $R_\tri$ is an $(n+1)$-ary relation on $W$.
\end{definition}

\begin{definition}[$\modsim$-model]
\label{def:models}
    Given a modal language $\modL{\tau, \Prop}$ and a $\modsim$-frame $\fr$, a $\modsim$-model $\model$ based on $\fr$ is a tuple $\model = \langle\fr, V \rangle$, where $V: \Prop \longrightarrow \mathcal{P}(w)$ is a \emph{valuation function}, assigning to every propositional letter $p$ the set of worlds at which $p$ is true. 
\end{definition}

\begin{definition}
\label{def:satisfiability}
    Given a modal language $\modL{\tau, \Prop}$,  let $\model $ be a $\modsim$-model and $w$ be a world in it. For any formula $ A $ of $\modL{\tau, \Prop}$, we define the \emph{satisfiability} relation of $\varphi$ at $\model, w$ inductively as follows:
    \begin{itemize}
        \item $\model, w \models p$ iff $w \in V(p)$; 
        \item $\model, w \not \models \bot$;
        \item $\model, w \models \NEG \varphi$ iff $\model, w \not \models \varphi$; 
        \item $\model, w \models \varphi \OR \psi$ iff $\model, w \models \varphi$ or $\model, w \models \psi$; 
        \item $\model, w \models \tri (\varphi_1, \dots,\varphi_n)$, for $n >0$, iff for some $v_1, \dots, v_n$ in $W$ such that $R_\tri(w, v_1, \dots, v_n)$, it holds that $\model, v_i \models \varphi_i$, for each $i \leq n$; 
        \item $\model, w \models \tri (\varphi_1, \dots,\varphi_n)$, for $n =0$, iff $w \in R_\tri$.
    \end{itemize}
\end{definition}

The following notions, standard in modal logic, generalise the notion of satisfiability to models, frames and classes of frames. 

\begin{definition}
\label{def:sat:true:valid}
    Let  $\varphi$ and $\psi$ be formulas of a modal language $\modL{\modsim, \Prop}$. We say that: 
    \begin{itemize}
        \item $\varphi$ is \emph{satisfiable} iff there is a $\modsim$-model $\model$ and a world $w$ in it such that $\model, w \models \varphi$;
        \item $\varphi$ is \emph{true} at a $\modsim$-model $\model$, in symbols $\model \models \varphi$, iff for every $w$ in $\model$, it holds that $\model, w \models \varphi$. 
        \item $\varphi$ is \emph{valid at a world $w$ of a $\modsim$-frame $\fr$} iff for every model $\model$ based on $\fr$, it holds that $\model, w \models \varphi$; 
		\item $\varphi$ is \emph{valid at a $\modsim$-frame $\fr$}, notation $\fr \models \varphi$,   iff $\varphi$ is valid at every world in $W$.  
        \item $\varphi$ is \emph{valid in a class $\mathcal{C}$ of $\modsim$-frames}, in symbols $\models_\mathcal{C}A$, iff, for every $\modsim$-model $\model$ based on a $\modsim$-frame $\fr \in \mathcal{C}$, it holds that $\model \models \varphi$.
        \item $\varphi$ and $\psi$ are \emph{logically equivalent in a class $\mathcal{C}$} iff it holds that $\models_{\mathcal{C}} \varphi \leftrightarrow \psi$. 
    \end{itemize}
    
\end{definition}

The basic modal logic $\K$ is defined as the set of formulas in the basic modal language which are valid on the class of all frames. One such vaild formula is $\axK: \BOX (A \IMP B) \IMP (\BOX A \IMP \BOX B) $. Adding this formula and the inference rule of \emph{necessitation} (from $ A$ infer $\BOX A$) to an axiom system for classical propositional logic yields an axiom system for $\K$ (in a language where $\DIA$ is primitive, we also need $\DIA \varphi \leftrightarrow \lnot \BOX \lnot \varphi$). 

\begin{figure}
		\begin{center}
			
			\begin{tikzpicture}[scale=0.7]
				\node (K) at (-2,-2) {\textcolor{blue}{$\mathsf{K}$}};
				\node (T) at (-2,2) {\textcolor{blue}{$\mathsf{T}$}};
				\node (TB) at (3,2) {\textcolor{blue}{$\mathsf{TB}$}};
				\node (KB) at (3,-2) {\textcolor{blue}{$\mathsf{KB}$}};
				
				\node (K45) at (2,-1) {\textcolor{blue}{$\mathsf{K45}$}};
				\node (K5) at (0,-1.5) {\textcolor{blue}{$\mathsf{K5}$}};
				\node (D5) at (0,0.5) {\textcolor{blue}{$\mathsf{D5}$}};
				
				\node (D) at (-2,0) {\textcolor{blue}{$\mathsf{D}$}};
				\node (D4) at (-1,1) {\textcolor{blue}{$\mathsf{D4}$}};
				\node (DB) at (3,0) {\textcolor{blue}{$\mathsf{DB}$}};		
				\node (D45) at (2,1) {\textcolor{blue}{$\mathsf{D45}$}};
				
				\node (S4) at (-1,3)  {\textcolor{blue}{$\mathsf{S4}$}};
				\node (S5) at (4,3) {\textcolor{blue}{$\mathsf{S5}$}};
				\node (KB5) at (4,-1) {\textcolor{blue}{$\mathsf{KB5}$}};
				\node (K4) at (-1,-1)  {\textcolor{blue}{$\mathsf{K4}$}};

				\draw[-, dashed] (K) -- (K4);	
				\draw[-,dashed] (K4) -- (K45);
				\draw[-,dashed] (K45) -- (KB5);
				\draw[-] (KB5) -- (KB);
				\draw[-] (KB) -- (K);
				
				\draw[-] (K) -- (D);
				\draw[-] (D) -- (T);

				\draw[-] (KB) -- (DB);
				\draw[-] (DB) -- (TB);
				\draw[-] (D) -- (DB);

				\draw[-] (KB5) -- (S5);
				\draw[-,dashed] (K4) -- (D4);
				\draw[-,dashed] (D4) -- (S4);
				\draw[-] (T) -- (S4);
				\draw[-] (S4) -- (S5);
				\draw[-] (S5) -- (TB);
				\draw[-] (TB) -- (T);	
				
				\draw[-,dashed] (K) -- (K5);
				\draw[-,dashed] (K5) -- (K45);
				
				\draw[-,dashed] (D) -- (D4);
				\draw[-,dashed] (D4) -- (D45);
				\draw[-,dashed] (K45) -- (D45);
				\draw[-,dashed] (D) -- (D5);
				\draw[-,dashed] (D5) -- (D45);
				\draw[-,dashed] (K5) -- (D5);
				\draw (D45) edge[ out=25,in=250,-,dashed] (S5);	
			\end{tikzpicture}
		\end{center}
		\caption{The $\sfive$-cube of modal logic}
		\label{fig:cube}
	\end{figure}
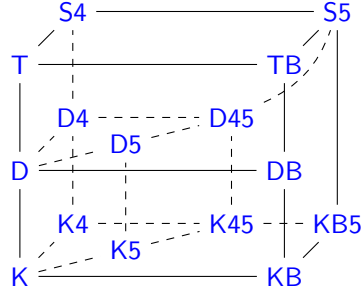

It is natural to require the accessibility relation of Kripke frames to satisfy additional properties, called \emph{frame conditions}. In this way, we can generate a number of modal logics. We specify below some well-known frame conditions, which generate the 15 logics in the so-called $\sfive$-cube of modal logic, depicted in \Cref{fig:cube}. 

	\begin{table}[h!]
		\begin{tabular}{c | l | l  l}
		Name & Axiom & \multicolumn{2}{l}{Frame  condition} \\
		\hline 
		&&& \\[-0.3cm]
		$ \axD $ & $ \Box A \IMP \Diamond A $ & Seriality & $ \forall x \exists y({\rel{x}{y}}) $ \\[0.05cm]
		$ \axT $ & $ \Box A \IMP A $ & Reflexivity & $ \forall x ({\rel{x}{x}}) $ \\[0.05cm]
		$ \axB $ & $ A \IMP \Box \Diamond A  $ & Symmetry& $ \forall x \forall y ({\rel{x}{y} } \, \Rightarrow \, {\rel{y}{x}}) $\\[0.05cm]
		4 & $ \Box A \IMP \Box \Box A  $& Transitivity & $ \forall x \forall y \forall z(({\rel{x}{y}} \, \& \, {\rel{y}{z}} ) \, \Rightarrow \, {\rel{x}{z}}) $\\[0.05cm]
		5 & $ \Diamond A \IMP \Box \Diamond A $& Euclideaness & $ \forall x \forall y \forall z (({\rel{x}{y}} \, \& \, {\rel{x}{z}} ) \, \Rightarrow \, {\rel{y}{z}}) $\\
		\end{tabular}
	\caption{Axioms and frame conditions}
	\label{tb:frames}
\end{table}

Every logic in  \Cref{fig:cube} is characterised by the class of frames denoted by the letters in the name of the logic. For instance, $\mathsf{D}$ is set of formulas valid at serial frames, and $\mathsf{KB5}$ is the set of formulas valid at symmetric and Euclidean frames. We have followed standard naming practice, and have not included `$\K$' in each name. Exceptions to the naming conventions are $\sfour$ and $\sfive$. Logic $\sfour$ is the set of all formulas valid at reflexive and transitive frames, while $\sfive$, the strongest logic in the family, is the set of all formulas valid at frames whose accessibility relation is an equivalence relation. So we can take $\sfour $ to be specified by $ \{ \axT, 4\} $ and $\sfive $ to be specified by $\{ \axT, 4, 5\} $, or equivalently by $ \{\axT, 4, \axB \}$. 

The formulas listed in the middle column of \Cref{tb:frames} are the axioms needed to be added to an axiom system for $\K$ to define axiom systems characterising the set of validities in the corresponding class of frames. For instance, an axiom system for $\mathsf{T}$ is defined by adding axiom $\axT$ to the axiom system for $\K$. Adding frame conditions restricts the class of models, and thus increases the set of valid propositions. Formally, we have the following result:

	\begin{theorem} 
    \label{thm:more:frames}
    Let $\modsim$ be a modal similarity type, $\mathcal{C}_1$ and $\mathcal{C}_2$ be two classes of $\modsim$-frames, and let $\logic_1$ and $\logic_2$ be the sets of formulas of that type valid at $\mathcal{C}_1$ and $\mathcal{C}_2$. 
		It holds that: 
		$$
		\mathcal{C}_1 \subseteq \mathcal{C}_2 \quad  \text{ iff } \quad  \logic_2 \subseteq \logic_1 \text{.}
		$$		
	\end{theorem}

There is a deeper link between the formulas from \Cref{tb:frames} and their frame conditions: each such formula \emph{defines} the corresponding class of frames, in the sense specified next. 
	
	\begin{definition}
    \label{def:mod:def}
    Let $\modsim$ be a modal similarity type, $\varphi$ a formula of this type, and $\cl$ be a class of $\modsim$-frames. We say that $A$ \emph{defines} $\mathcal{C}$ if, for all frames $\fr$ we have that:
		$$
			\label{eq:def}
		\fr \in \mathcal{C} \quad  \text{ iff } \quad \fr \models \varphi
		$$
		If there is a formula $\varphi$ such that $\varphi$ defines $\mathcal{C}$, we say that $\mathcal{C}$ is \emph{modally definable}. 
	\end{definition}

We only considered simple frame conditions in the simpler modal language; many more meaningful frame conditions have been defined in the literature. For instance, we will later consider frames with the \emph{diamond property}, which expresses a form of confluence: 
\[
\forall x\forall y\forall z ((xRy \mathbin\& xRz) \Rightarrow \exists w (yRw \mathbin\& zRw)) \text{.}\]

\section{A Generic Modal Framework}\label{sec:framework}

In this section, we present our generic framework for modal languages and reasoning.
In later sections, when it is necessary to distinguish definitions or theorems in this generic development, we refer to the namespace \lstinline|Modal|.

\subsection{Operator Signatures}
\label{sec:polynomial-signatures}

We start our development by generalising modal similarity types so that operators are mapped to generic types (instead of finite arities).
That is, a modal similarity type \lstinline|τ| consists of a type \lstinline|A : Type uA| and a mapping \lstinline|B : A -> Type uB| for some universes \lstinline|uA| and \lstinline|uB|. For any operator \lstinline|a : τ.A|, \lstinline|τ.B a| is the type of its argument positions.

For the present development, the main advantages of this design choice are:
\begin{itemize}
\item It keeps our development abstract. We do not require (i) that argument positions are natural numbers, (ii) that argument positions can be ordered, nor (iii) that operators have finite arities.
\item It enables reuse. This structure is exactly a \emph{polynomial functor} \cite{NS25}, as found in Mathlib. This makes some of our definitions (e.g., about unary polynomial functors, shown later) useful to the general Lean ecosystem instead of just modal logic.
\end{itemize}

Formally, a modal similarity type \lstinline|τ| has type \lstinline|PFunctor|, from Mathlib:
\begin{lstlisting}
structure PFunctor.{uA,uB} where
  A : Type uA
  B : A -> Type uB
\end{lstlisting}
We call these also operator signatures, or just signatures.

\subsection{Syntax}

The syntax of propositions is modelled as an inductive data type parametrised over a signature \lstinline|τ| and a type of atomic propositions \lstinline|Atom|.\footnote{In \cslib logics, the convention is that formulas are called propositions, so we adopt this terminology.}
\begin{lstlisting}
inductive Proposition (τ : PFunctor) Atom where
  | atom (p : Atom) | false | not (φ : Proposition τ Atom)
  | or (φ₁ φ₂ : Proposition τ Atom)
  | triangle (op : τ.A) (φs : τ.B op → Proposition τ Atom)
\end{lstlisting}
This formalises a generalisation of the syntax given in \Cref{sec:background:modal}, with the case of triangle $\tri$ adapted to polynomial functors. Our language is more general because of the reasons connected to \lstinline|PFunctor| explained in \Cref{sec:polynomial-signatures} and the fact that we impose no requirements on \lstinline|Atom| (notably, we do not need it to be countable).

To obtain the usual notation, we first instantiate the existing notation typeclasses for logical operators that match our constructors. The next snippet is the instance for `\lstinline|¬|'.
\begin{lstlisting}
instance {τ : PFunctor} {Atom : Type*} : HasNot (Proposition τ Atom) := ⟨.not⟩
\end{lstlisting}
For the modal operator \lstinline|Δ|, we add a new notation typeclass to \cslib.
\begin{lstlisting}
(*@\lstlabel{HasTriangle}@*)class HasTriangle (α : Type*) (τ : outParam PFunctor) where
  triangle (op : τ.A) (arg : τ.B op → α) : α
\end{lstlisting}
This allows us to write \lstinline|Δ[op]φs| for the application of the modal operator \lstinline|op|, where \lstinline|φs| is the argument -- a map from argument positions to propositions.

\newcommand{\defeql}{\mathbin{\texttt{:=}}}

We then derive the other propositional operators as usual, by defining them as follows through appropriate instances of operator typeclasses.
\begin{align}
\top &\defeql \neg\bot \label{eq:top-def} \\
\varphi_1 \land \varphi_2 &\defeql \neg(\neg\varphi_1 \lor \neg\varphi_2) \label{eq:and-def} \\
\varphi_1 \to \varphi_2 &\defeql \neg\varphi_1 \lor \varphi_2 \label{eq:imp-def} \\
\varphi_1 \leftrightarrow \varphi_2 &\defeql (\varphi_1 \to \varphi_2) \land (\varphi_2 \to \varphi_1) \label{eq:iff-def}
\end{align}

It is useful at this point to introduce an abbreviation for the proposition maps used as arguments for modal operators and define APIs for it.
\begin{lstlisting}
abbrev PropositionMap τ op Atom := τ.B op → Proposition τ Atom
\end{lstlisting}
We define the expected point-wise extensions of all propositional connectives to proposition maps, for example:
\begin{lstlisting}[end={.}]
def PropositionMap.or (φs₁ φs₂ : PropositionMap τ op Atom) :=
  fun i => φs₁ i ∨ φs₂ i
\end{lstlisting}
Again, we instantiate all appropriate typeclasses from \cslib to obtain notation for proposition maps, so that we can write \lstinline|¬φs|, \lstinline|φs₁ ∨ φs₂|, etc., in Lean.

With this setup, we can now define a notation typeclass for nabla $\trid$ (\lstinline|HasNabla|) and derive this operator as follows.
\begin{equation}
\nabla\texttt{[op]}\varphi\texttt{s} \defeql \neg\Delta\texttt{[op]}\neg\varphi\texttt{s} \label{eq:nabla-def}
\end{equation}

\subsection{Semantics}\label{sec:semantics}

Moving to semantics, we first generalise the notion of $\tau$-frame from \Cref{def:frames} to our operator signatures:
\begin{lstlisting}[end={.}]
structure Frame (World : Type u) (τ : PFunctor) where
  r : (op : τ.A) → World → (τ.B op → World) → Prop
\end{lstlisting}
A model is then a frame extended with a valuation, as detailed in \Cref{def:models},
\begin{lstlisting}[end={,}]
structure Model (World : Type u) (τ : PFunctor) (Atom : Type v)
    extends Frame World τ where
  v : World → Atom → Prop
\end{lstlisting}
which is enough to define satisfaction (\Cref{def:satisfiability}): 
\begin{lstlisting}[end={.}]
(*@\lstlabel{Satisfies}@*)def Satisfies (m : Model World τ Atom) (w : World) :
    Proposition τ Atom → Prop
  | .atom p => m.v w p
  | .false => False
  | .not φ => ¬Satisfies m w φ
  | .or φ₁ φ₂ => Satisfies m w φ₁ ∨ Satisfies m w φ₂
  | .triangle op φs => ∃ ws : τ.B op → World, m.r op w ws ∧ ∀ i, Satisfies m (ws i) (φs i)
\end{lstlisting}

We now enter the semantic infrastructure of \cslib for this kind of systems. First, we define a data structure for satisfaction judgements, called \lstinline|Judgement|, that bundles (contains) a model (\lstinline|m|), a world (\lstinline|w|), and a proposition \lstinline|φ|. We write \lstinline|Modal[m,w ⊨ φ]| for such a judgement.
We then instantiate \cslib's \lstinline|InferenceSystem| class to elicit that \lstref{Satisfies} is the canonical way of determining the validity of such judgements:\footnote{Saying that \lstref{Satisfies} is an `inference system' might surprise some readers. This should not be confused with defining a sequent calculus. In particular, \cslib intends inference system quite broadly as a means to derive a statement (even semantic ones), a practice used broadly in theory of programming languages \cite{H16,M23}.}
\begin{lstlisting}[end={,}]
instance {World : Type*} {τ : PFunctor} {Atom : Type*} :
    HasInferenceSystem (Judgement World τ Atom) := ⟨Satisfies.Bundled⟩
\end{lstlisting}
where \lstinline|Satisfies.Bundled| just applies \lstref{Satisfies} to the components of the judgement.
Instantiating \lstinline|InferenceSystem| gives access to the notation \lstinline|⇓Modal[m,w ⊨ φ]|, read `a witness of \lstinline|Modal[m,w ⊨ φ]|'. Thus, concretely, this expands to \lstinline|Satisfies m w φ|.

With the basic semantic infrastructure in place, we can start proving theorems about the semantics of operators. The first interesting ones prove the correctness of our derived operators -- \Cref{eq:top-def,eq:and-def,eq:imp-def,eq:iff-def,eq:nabla-def} -- by showing that they exhibit the expected semantics. We give some examples next.
\begin{lstlisting}
(*@\lstlabel{Satisfies.and_iff_and}@*)theorem Satisfies.and_iff_and {m : Model World τ Atom} :
    ⇓Modal[m,w ⊨ φ₁ ∧ φ₂] ↔ ⇓Modal[m,w ⊨ φ₁] ∧ ⇓Modal[m,w ⊨ φ₂]
theorem Satisfies.nabla_iff_forall {m : Model World τ Atom} :
    ⇓Modal[m,w ⊨ ∇[op]φs] ↔
      ∀ ws, m.r op w ws → ∃ i, ⇓Modal[m,(ws i) ⊨ (φs i)]
(*@\lstlabel{Satisfies.imp_iff_imp}@*)theorem Satisfies.imp_iff_imp {m : Model World τ Atom} :
    ⇓Modal[m,w ⊨ φ₁ → φ₂] ↔ (⇓Modal[m,w ⊨ φ₁] → ⇓Modal[m,w ⊨ φ₂])
(*@\lstlabel{Satisfies.iff_iff_iff}@*)theorem Satisfies.iff_iff_iff {m : Model World τ Atom} :
    ⇓Modal[m,w ⊨ φ₁ ↔ φ₂] ↔ (⇓Modal[m,w ⊨ φ₁] ↔ ⇓Modal[m,w ⊨ φ₂])
\end{lstlisting}

Satisfaction gives rise to two notions that will be significant later on.

One is the denotation of a proposition, i.e., its truth set \cite{BRV01,GNO21}.
\begin{lstlisting}
def Proposition.denotation (m : Model World τ Atom) :
    Proposition τ Atom → Set World
  | .atom p => {w | m.v w p}
  | .false => ∅
  | .not φ => (φ.denotation m)ᶜ
  | .or φ₁ φ₂ => φ₁.denotation m ∪ φ₂.denotation m
  | .triangle op φs =>
    {w | ∃ ws, m.r op w ws ∧ ∀ i, ws i ∈ (φs i).denotation m}
\end{lstlisting}
Denotations are linked to satisfaction by the following theorem, which will be important to prove the relationship between denotations and logical equivalence.
\begin{lstlisting}
(*@\lstlabel{satisfies_mem_denotation}@*)theorem satisfies_mem_denotation {m : Model World τ Atom}
    {φ : Proposition τ Atom} : w ∈ φ.denotation m ↔ ⇓Modal[m,w ⊨ φ]
\end{lstlisting}

The other notion derived from satisfaction is the \emph{theory} of a world under a model, i.e., the set of all propositions that it satisfies:
\begin{lstlisting}[end={.}]
abbrev theory {World : Type*} {τ : PFunctor} {Atom : Type*}
    (m : Model World τ Atom) (w : World) : Set (Proposition τ Atom) :=
  {φ | ⇓Modal[m,w ⊨ φ]}
\end{lstlisting}
We say that two worlds are \emph{theory equivalent} if they have the same theory.
\begin{lstlisting}
abbrev TheoryEq (m : Model World τ Atom) (w₁ w₂ : World) :=
  theory m w₁ = theory m w₂
\end{lstlisting}
We will come back to this concept in our development of Hennessy--Milner logic in \Cref{sec:hml}.

\section{Reusable Metatheory and Automation}
\label{sec:metatheory}

The definitions of the previous section are intentionally independent of any particular modal signature or class of frames. We exploit this generality to develop metatheory that is inherited by every subsequent specialisation.

This is an important part of our library-oriented approach. Reuse should not stop at syntax and semantics: notions such as axioms, logical equivalence, and routine semantic reasoning should also be developed once, against the generic interfaces. In later sections, we will show how this infrastructure can be put to work.

\subsection{Axioms and Frame Conditions}\label{sec:axioms}

A modal axiom expresses validity independently of a particular valuation. In the study of such axioms, the assumptions one makes on frames -- the \emph{frame conditions} -- are particularly important. We treat this as a first-class concern in our library.

We instantiate \lstinline|InferenceSystem| so that, for any frame \lstinline|f|, \lstinline|Axiom f ⇓φ| means that \lstinline|φ| is satisfied under \lstinline|f| for any valuation and world: 
\begin{lstlisting}[end={.}]
instance {World : Type*} {τ : PFunctor} {Atom : Type*} (f : Frame World τ) :
    InferenceSystem (Axiom f) (Proposition τ Atom) where
  derivation φ := ∀ v w, ⇓Modal[⟨f,v⟩,w ⊨ φ]
\end{lstlisting}
In other words, \lstinline|Axiom f ⇓φ| is a witness that \lstinline|φ| is valid for the frame \lstinline|f| (formalising the notation $\fr \models \varphi$ from \Cref{def:sat:true:valid} in \Cref{sec:background}).

We develop a database of general axioms, for example the generalised versions of the notable axiom K and duality:
\begin{lstlisting}[end={,}]
theorem Satisfies.k (f : Frame World τ) {φs : PropositionMap τ op Atom}
    [DecidableEq (τ.B op)] {i : τ.B op} {φ₁ φ₂ : Proposition τ Atom}
    (hi : φs i = (φ₁ → φ₂)) :
    Axiom f ⇓(∇[op]φs → (∇[op]φs[i := φ₁] → ∇[op]φs[i := φ₂]))

(*@\lstlabel{Satisfies.dual}@*)theorem Satisfies.dual (f : Frame World τ) {φs : PropositionMap τ op Atom} :
    Axiom f ⇓(Δ[op]φs ↔ ¬∇[op]¬φs)
\end{lstlisting}
where \lstinline|φs[i := φ]| stands for function update, i.e., \lstinline|φs| where \lstinline|i| is now mapped to \lstinline|φ|. The use of function updates is why we need the assumption that equality of the argument positions is decidable -- \lstinline|DecidableEq (τ.B op)|.

More axioms become provable under certain frame conditions, but modal logic is well known for yielding characterisation results on frame classes: some axioms are valid if and only if a frame has a certain property.
Consider, for example, axiom $\axT$ of basic modal logic (the modal logic with a single modal operator that takes exactly one argument, \lstinline|◇|): \lstinline|φ → ◇φ|; this is valid if and only if the binary relation associated to the \lstinline|◇| operator is reflexive.
In general modal logic, where frames can have multiple polyadic relations, we need to formulate appropriate generalisations of these frame conditions in order corresponding generalised versions of these axioms.

Let the \lstinline|diagonal| of a frame \lstinline|f| on an operator \lstinline|op| be the binary relation induced by restricting the \lstinline|op|'s accessibility relation to constant maps:
\begin{lstlisting}[end={.}]
def Frame.diagonal (f : Frame World τ) (op : τ.A) : World → World → Prop :=
  fun w w' => f.r op w (fun _ => w')
\end{lstlisting}
Similarly, let \lstinline|PropositionMap.const op φ| map every argument for \lstinline|op| to \lstinline|φ|.
We can then state the generalised version of axiom T for an operator \lstinline|op| and prove that it holds if and only if the diagonal on \lstinline|op| is reflexive.
\begin{lstlisting}
theorem Satisfies.t (f : Frame World τ) [instRefl : Std.Refl (f.diagonal op)]
    (φ : Proposition τ Atom) :
    Axiom f ⇓(φ → Δ[op](PropositionMap.const op φ))
\end{lstlisting}
\begin{lstlisting}
theorem Satisfies.t_refl (f : Frame World τ) {op : τ.A} [Nonempty Atom]
    (h : ∀ φ : Proposition τ Atom,
      Axiom f ⇓(φ → Δ[op](PropositionMap.const op φ))) :
    Std.Refl (f.diagonal op)
\end{lstlisting}
The proof of the latter uses an atom to instantiate \lstinline|φ| in \lstinline|h|, hence the requirement that \lstinline|Atom| be nonempty.

We build similar generalisations for other frame conditions, and prove that they correspond exactly to general versions of the axioms $\axB$, $4$, and $5$. We list them in order.
\begin{itemize}
\item A frame is diagonally symmetric on \lstinline|op| if whenever \lstinline|f.r op w ws|, some element of \lstinline|ws| is related to the constant world map for \lstinline|w|.
\begin{lstlisting}
class Frame.DiagonalSymm (f : Frame World τ) (op : τ.A) where
  symm w ws : f.r op w ws → ∃ i, f.r op (ws i) (fun _ => w)
\end{lstlisting}
\item A frame is transitive at \lstinline|op| if its accessibility relation can be composed through any accessible component.
\begin{lstlisting}
class Trans (f : Frame World τ) (op : τ.A) where
  trans w ws₁ i ws₂ : f.r op w ws₁ → f.r op (ws₁ i) ws₂ → f.r op w ws₂
\end{lstlisting}
\item A frame is right Euclidean at \lstinline|op| if, whenever a world accesses two argument maps, some component of the second accesses the first map.
\begin{lstlisting}
class RightEuclidean (f : Frame World τ) (op : τ.A) where
  rightEuclidean w ws₁ ws₂ :
    f.r op w ws₁ → f.r op w ws₂ → ∃ i, f.r op (ws₂ i) ws₁
\end{lstlisting}
\end{itemize}
These conditions correspond to the next axioms.
\begin{description}
\item[$\axB$:] \lstinline|φ → ∇[op](PropositionMap.const op (Δ[op](PropositionMap.const op φ)))|.
\item[$4$:] \lstinline|Δ[op]φs → Δ[op]ψs|, if there is \lstinline|i| such that \lstinline|φs i = (Δ[op]ψs)|.
\item[$5$:] \lstinline|Δ[op]φs → ∇[op](PropositionMap.const op (Δ[op]φs))|.
\end{description}

At the level of model classes, we can prove the expected underlying order-theoretic principle (formalising \Cref{thm:more:frames} from \Cref{sec:background}): restricting the class of models under consideration can only increase the set of valid propositions. 
Formally, let a proposition be \lstinline|valid| for a model class (given as a set) if it is satisfied for every model in the class at every world:
\begin{lstlisting}[end={.}]
def Proposition.valid {World : Type*} {τ : PFunctor} {Atom : Type*}
    (S : Set (Model World τ Atom)) (φ : Proposition τ Atom) : Prop :=
  ∀ m ∈ S, ∀ (w : World), ⇓Modal[m,w ⊨ φ]
\end{lstlisting}
Also, let the \lstinline|logic| of a model class be the set of all valid propositions for that class:
\begin{lstlisting}[end={.}]
def logic {World : Type*} {τ : PFunctor} {Atom : Type*}
    (S : Set (Model World τ Atom)) : Set (Proposition τ Atom) :=
  {φ | φ.valid S}
\end{lstlisting}
Then, \lstinline|logic| is antitone: it grows as its input model class decreases. In the formal statement, we can reuse the \lstinline|Antitone| definition from Mathlib:
\begin{lstlisting}[end={.}]
theorem logic_antitone {World : Type*} {τ : PFunctor} {Atom : Type*} :
    Antitone (logic (World := World) (τ := τ) (Atom := Atom))
\end{lstlisting}

\subsection{Logical Equivalence, Contexts, and Congruences}
\label{sec:logical-equivalence}

Two propositions are said to be \emph{logically equivalent} if they can be used interchangeably. This section is dedicated to formalising logical equivalence, its properties, and notable general laws.

\subsubsection{Logical Equivalence Under a Model}\label{sec:equiv-model}
We start by formalising equivalence under a particular model, which is straightforward:
\begin{lstlisting}[end={.}]
def Proposition.Equiv (m : Model World τ Atom) (φ₁ φ₂ : Proposition τ Atom) :
    Prop := ∀ (w : World), ⇓Modal[m,w ⊨ φ₁ ↔ φ₂]
\end{lstlisting}
Yet this is still just a definition, which we want to prove adequate.
Specifically, we are interested in proving that the relation is a congruence (it is closed under all syntactic contexts) that preserves satisfaction.

Showing that logical equivalence preserves satisfaction is straightforward, thanks to theorem \lstref{Satisfies.iff_iff_iff} from \Cref{sec:semantics}; in particular, \lstinline|⇓Modal[m,w ⊨ φ₁ ↔ φ₂]| implies \lstinline|⇓Modal[m,w ⊨ φ₁] ↔ ⇓Modal[m,w ⊨ φ₂]|.
By contrast, formalising that logical equivalence is a congruence requires more work, which we discuss in the rest of this section.

First, we instantiate the congruence notation typeclass for our equivalence relation:
\begin{lstlisting}[end={.}]
instance : Congruence (Proposition.Equiv m) := ⟨⟩
\end{lstlisting}
This gives access to the notation \lstinline|φ₁ ≡[Equiv m] φ₂| for \lstinline|Proposition.Equiv m φ₁ φ₂|.

We then prove that, for any model \lstinline|m|, \lstinline|≡[Equiv m]| is an equivalence relation, i.e., it is reflexive, symmetric, and transitive.
\begin{lstlisting}
instance (m : Model World τ Atom) :
    IsEquiv (Proposition τ Atom) (Proposition.Equiv m)
\end{lstlisting}

Next, we have to prove that \lstinline|≡[Equiv m]| is, in \cslib's terminology, a \emph{lawful} congruence; that is, it is closed under all syntactic contexts for propositions.
Informally, a propositional context is a proposition with a single `hole', which can later be filled with another proposition. For example, \lstinline|φ₁ ∨ ·| is a context, and filling it with \lstinline|φ₂| returns \lstinline|φ₁ ∨ φ₂|.

We formalise propositional contexts as follows.
\begin{lstlisting}[end={,}]
(*@\lstlabel{Proposition.Context}@*)inductive Proposition.Context (τ : PFunctor) (Atom : Type u) where
  | hole
  | not (c : Context τ Atom)
  | orL (c : Context τ Atom) (φ : Proposition τ Atom)
  | orR (φ : Proposition τ Atom) (c : Context τ Atom)
  | triangle (op : τ.A) (i : τ.B op) (c : Context τ Atom)
    (φs : PropositionMap.Without τ Atom op i)
\end{lstlisting}
where \lstinline|PropositionMap.Without| is a proposition map missing an argument at some position \lstinline|i|,
\begin{lstlisting}[end={.}]
abbrev PropositionMap.Without (τ : PFunctor) (Atom : Type*) (op : τ.A)
    (i : τ.B op) := {j : τ.B op // j ≠ i} → Proposition τ Atom
\end{lstlisting}
Filling a context is a simple recursive function.
\begin{lstlisting}
def Proposition.Context.fill {τ : PFunctor} [τ.DecidableEqChildren]
    {Atom : Type*} (c : Context τ Atom) (φ : Proposition τ Atom) :=
  match c with
  | hole => φ | not c => .not (c.fill φ)
  | orL c φ' => (c.fill φ).or φ' | orR φ' c => φ'.or (c.fill φ)
  | .triangle op i c φs => .triangle op fun j =>
    if h : j = i then c.fill φ else φs ⟨j, h⟩
\end{lstlisting}
The case for triangle is what requires knowing that argument positions have decidable equality, i.e., \lstinline|τ.DecidableEqChildren|.
\lstinline|DecidableEqChildren| is a new general typeclass that we contribute to \lstinline|PFunctor| in \cslib, since it abstracts from the fact that the polynomial functor is used to represent a modal operator signature.

With these definitions in place, we can register \lstref{Proposition.Context} as the canonical notion of syntactic context for modal propositions in \cslib's infrastructure:
\begin{lstlisting}[end={.}]
instance {τ : PFunctor} [τ.DecidableEqChildren] {Atom : Type*} :
    HasContext (Proposition τ Atom) := ⟨Proposition.Context.fill⟩
\end{lstlisting}
This permits writing, for any context \lstinline|c| and proposition \lstinline|φ|, \lstinline|c<[φ]| for \lstinline|c.fill φ|.

We can finally prove that our logical equivalence relation is a lawful congruence according to \cslib's definition:
\begin{lstlisting}[end={.}]
instance {τ : PFunctor} [τ.DecidableEqChildren] {Atom : Type*}
    (m : Model World τ Atom) : LawfulCongruence (Proposition.Equiv m)
\end{lstlisting}
This requires proving that the next statement holds for any world \lstinline|w|, context \lstinline|c|, and propositions \lstinline|φ₁| and \lstinline|φ₂|.
\begin{lstlisting}[end={.}]
 ⇓Modal[m,w ⊨ c<[φ₁] ↔ c<[φ₂]]
\end{lstlisting}
We proceed by structural induction on \lstinline|c|.
The proof benefits from a careful strategy.
Bi-implication is a derived connective, so if we naively followed standard simplification strategies and unfolded it to primitive operators (recall \Cref{eq:top-def,eq:and-def,eq:imp-def,eq:iff-def,eq:nabla-def}) we would get cumbersome shapes. For example, the case in which \lstinline|c| is \lstinline|φ ∨ ·| would yield the goal
\begin{lstlisting}[end={.}]
 ⇓Modal[m,w ⊨ ¬(¬(¬(φ ∨ φ₁) ∨ (φ ∨ φ₂)) ∨ ¬(¬(φ ∨ φ₂) ∨ (φ ∨ φ₁)))]
\end{lstlisting}
Instead, we can direct the proof to use our characterisation theorems that give a semantics to derived operators, like \lstref{Satisfies.iff_iff_iff} from \Cref{sec:semantics}.
With that, we quickly get to two very easy subgoals: \lstinline|⇓Modal[m,w ⊨ φ ∨ φ₁]|, knowing that \lstinline|⇓Modal[m,w ⊨ φ ∨ φ₂]|; and, conversely, \lstinline|⇓Modal[m,w ⊨ φ ∨ φ₂]|, knowing that \lstinline|⇓Modal[m,w ⊨ φ ∨ φ₁]|.
We will come back to how strategies like this can be automated and made generally available in \Cref{sec:automated-modal-reasoning}.

We conclude this part by proving that logical equivalence under a model admits an equivalent extensional characterisation: two propositions are equivalent exactly when they have the same denotation.
\begin{lstlisting}
theorem Proposition.equiv_iff_denotation_eq :
    (φ₁ ≡[Equiv m] φ₂) ↔ φ₁.denotation m = φ₂.denotation m
\end{lstlisting}
The proof of this result uses theorem \lstref{satisfies_mem_denotation} from \Cref{sec:semantics}.

\subsubsection{Equivalence Within a Model Class and Universal Equivalence}

Building on our model-dependent logical equivalence, we can introduce logical equivalence within a model class \lstinline|S|, written \lstinline|≡[EquivWithin S]|. It states that two propositions are logically equivalent within \lstinline|S|:
\begin{lstlisting}[end={.}]
def Proposition.EquivWithin (S : ModelClass World τ Atom)
    (φ₁ φ₂ : Proposition τ Atom) : Prop := ∀ m ∈ S, φ₁ ≡[Equiv m] φ₂
\end{lstlisting}
This relation directly implies co-validity in the same class:
\begin{lstlisting}[end={.}]
theorem Proposition.equivWithin_valid (S : ModelClass World τ Atom)
    (φ₁ φ₂ : Proposition τ Atom) (h : φ₁ ≡[EquivWithin S] φ₂) :
    (φ₁.valid S ↔ φ₂.valid S)
\end{lstlisting}
The structural properties established in \Cref{sec:equiv-model} lift pointwise to model classes.
In particular, \lstinline|EquivWithin S| is again an equivalence relation and,
under the same decidability assumption on argument positions, a lawful
congruence. Thus replacing logically equivalent propositions inside arbitrary
modal contexts is valid uniformly throughout the class.

By fixing the argument \lstinline|S| of \lstinline|≡[EquivWithin S]|, we obtain
notions of logical equivalence for different logics, represented by different
classes of models (for example, those induced by different frame conditions).

A particularly useful instance is obtained by taking \lstinline|S| to be the
universal set of models (relative to a fixed world type), \lstinline|Set.univ|.
We call the resulting relation \emph{universal equivalence}, written \lstinline|≡[UEquiv]|:
\begin{lstlisting}[end={.}]
abbrev Proposition.UEquiv {World : Type*} (φ₁ φ₂ : Proposition τ Atom) :=
  EquivWithin (World := World) Set.univ φ₁ φ₂
\end{lstlisting}
Since it is a special case of \lstinline|EquivWithin|, it inherits all its properties.

Universal equivalence is the strongest of the three notions for a fixed type of worlds: equivalent propositions must agree in every model, rather than in one particular model or model class.
This makes it a good candidate for automation, as we discuss in \Cref{sec:automation:equiv}.

\subsection{Automated Modal Reasoning}\label{sec:automated-modal-reasoning}

The generic theory developed so far is intended to be reused across the modal logics built on our framework. We leverage this reuse to develop automation for routine reasoning that applies uniformly across those logics.
The idea is that reuse should extend beyond definitions and theorems: common proof patterns should also be captured once and inherited by subsequent specialisations.

To this end, we equip the development with a dedicated database for Lean's
\lstinline|grind| tactic. \lstinline|grind| is Lean's extensible automated reasoning tactic, which combines simplification with pattern-based theorem instantiation using E-matching \cite{MM26}.
Our database, called \lstinline|modal|, collects carefully selected rules for modal reasoning. It can be added to the usual
\lstinline|grind| invocation with
\begin{lstlisting}[end{,}]
grind [modal]
\end{lstlisting}
which runs it together with all \lstinline|grind|-annotated theorems in scope, or
used in isolation with
\begin{lstlisting}
grind only [modal]
\end{lstlisting}
when one wants to restrict automation to the rules exposed by the modal
development. The latter form can be useful for performance and, also, for checking that a proof relies purely on modal reasoning.

The \lstinline|modal| database is extensible: as we derive more specialised
modal interfaces, we add the corresponding reasoning rules to the same
database. Thus, later developments can invoke the same automation interface
while reasoning directly with unary or unimodal operators, rather than
manually reducing them to the underlying generic polyadic representation. This is going to be very important in \Cref{sec:derived-modal-operators,sec:deriving-modal-logics}.

\subsubsection{Semantic Characterisation Rules}

We begin the development of the \lstinline|modal| database by adding semantic characterisation theorems for primitive and derived logical operators.
These allow \lstinline|grind| to decompose and reason about propositions without unfolding their definitions or the satisfaction relation directly.

For example, the following proof combines the semantic characterisations of
conjunction and implication (\lstref{Satisfies.and_iff_and} and \lstref{Satisfies.imp_iff_imp} from \Cref{sec:semantics}) to automatically derive modus ponens:
\begin{lstlisting}[end={.}]
example (h : ⇓Modal[m,w ⊨ (φ → ψ) ∧ φ]) : ⇓Modal[m,w ⊨ ψ] := by
  grind only [modal]
\end{lstlisting}
Inspecting the rules used by \lstinline|grind| (which can be done by invoking it with \lstinline|grind?|) reveals precisely the two
relevant characterisation theorems:
\begin{lstlisting}
grind only [= Satisfies.and_iff_and, = Satisfies.imp_iff_imp]
\end{lstlisting}
This illustrates that, although conjunction and implication are derived operators in our syntax, proofs can treat them according to their usual semantics. The details of their encoding in terms of the primitive constructors are therefore hidden from routine reasoning.

Not every semantic characterisation theorem is included in the
\lstinline|modal| database: we deliberately exclude the
characterisations of \lstinline|Δ| and \lstinline|∇|. Their semantics exposes the accessibility relations and introduces quantification over maps of successor worlds, which can considerably enlarge the search space explored by \lstinline|grind|. Since many proofs do not need to inspect the accessibility relation at all, enabling these rules by default would make routine reasoning unnecessarily expensive.

When a proof does need to reason directly about the semantics of the modal
operators, the corresponding characterisation theorem can simply be supplied
explicitly, for example,
\begin{lstlisting}[end={.}]
grind only [modal, Satisfies.triangle_iff_exists, Satisfies.nabla_iff_forall]
\end{lstlisting}
Thus, the \lstinline|modal| database provides a useful default level of
automation without preventing users from exposing lower-level semantic
structure when needed.

\subsubsection{Axioms and Equivalences}\label{sec:automation:equiv}
The \lstinline|modal| database also contains rules that connect the different
layers of our metatheory. One important example concerns axioms and logical
equivalence. Recall that an axiom for a frame is valid under every valuation,
whereas logical equivalence under a model fixes a particular valuation. These
notions are related by the following characterisation:
\begin{lstlisting}[end={.}]
theorem Proposition.axiom_iff_forall_equiv
    (f : Frame World τ) (φ₁ φ₂ : Proposition τ Atom) :
    (Axiom f ⇓(φ₁ ↔ φ₂)) ↔ ∀ v, φ₁ ≡[Equiv ⟨f,v⟩] φ₂
\end{lstlisting}
We include this theorem in our \lstinline|modal| database, so that any bi-implication axiom can be used directly as a logical equivalence under any valuation.
Combined with the definitions of equivalence over model classes, this allows general modal axioms to be turned into reusable equivalences with little proof overhead.

We also include theorems connecting our three notions of logical equivalence -- under a model, within a model class, and universally -- so that grind can move between these levels when the required hypotheses are available.
For an example that traverses the entire hierarchy, recall \lstref{Satisfies.dual} from \Cref{sec:axioms}, which concludes:
\begin{lstlisting}[end={.}]
Axiom f ⇓(Δ[op]φs ↔ ¬∇[op]¬φs)
\end{lstlisting}
With \lstinline|modal|, we can automatically lift this axiom to a universal equivalence:
\begin{lstlisting}[end={.}]
theorem Proposition.dual_equiv (φs : PropositionMap τ op Atom) :
    Δ[op]φs ≡[UEquiv (World := World)] ¬∇[op]¬φs := by
  grind only [modal, Satisfies.dual]
\end{lstlisting}
Inspecting the \lstinline|grind| call reveals that the proof uses exactly our \lstinline|modal| theorems for moving an axiom with a bi-implication to a model-specific equivalence, and then equivalence for all models in a class to class-specific equivalence (in this case the universal set).

\section{Derived Modal Interfaces}
\label{sec:derived-modal-operators}

In this section, we derive higher-level interfaces from the generic polyadic framework developed so far. These interfaces serve two purposes. First, they recover the familiar unary and unimodal presentations used by standard modal logics. Second, they connect modal reasoning to ordinary Lean predicates, sets, and other container-like structures.

The central benefit is that neither kind of interface requires a new modal syntax or satisfaction relation. Instead, both are obtained from the generic structures and metatheory developed previously.
In the next section, we will use these interface to derive different modal logics, including the one that previously existed as a separate development in \cslib.

\subsection{Derived Notation}
\label{sec:derived-notation}
\cslib provides some notation typeclasses relevant to our development, which were originally introduced for ad-hoc developments of specialised modal logics (like Hennessy--Milner Logic and basic modal logic).
\begin{itemize}
\item \lstinline|HasDynamicDiamond|, which gives access to the notation \lstinline|d⟨op⟩φ|. This is usually written \lstinline|⟨op⟩φ| in the literature: the \lstinline|d| is an immaterial literal, used only to avoid conflicts with unrelated core Lean notation.
\item \lstinline|HasDynamicBox|, which gives access to the notation \lstinline|d[op]φ|. (Same comment for the \lstinline|d| as in the previous point.)
\item \lstinline|HasDiamond|, which gives access to the notation \lstinline|◇|.
\item \lstinline|HasBox|, which gives access to the notation \lstinline|□|.
\end{itemize}

Our development organises these classes in a clear hierarchy based on our new \lstref{HasTriangle} and \lstinline|HasNabla|.

Instances of \lstinline|HasDynamicDiamond| and \lstinline|HasDynamicBox| can always be automatically derived from the presence of, respectively, a \lstinline|Δ| or a \lstinline|∇| operator.
For example, when a \lstinline|Δ| operator is available, then we can write \lstinline|d⟨op⟩φ| for \lstinline|Δ[op](fun _ => φ)| (the application of \lstinline|Δ[op]| that supplies \lstinline|φ| for all arguments):
\begin{lstlisting}[end={.}]
instance [HasTriangle α τ] : HasDynamicDiamond α τ.A where
  dynDiamond op φ := Δ[op](PFunctor.const op φ)
\end{lstlisting}
Likewise for \lstinline|∇| and \lstinline|d[op]φ|:
\begin{lstlisting}
instance [HasNabla α τ] : HasDynamicBox α τ.A where
  dynBox op φ := ∇[op](PFunctor.const op φ)
\end{lstlisting}

Say now that that we have a language with a unique modal operator.
We model uniqueness with Mathlib's \lstinline|Unique|. Briefly, \lstinline{Unique} extends Lean's \lstinline{Inhabited}, which states that a type has a \lstinline|default| element, with the property that any element of the type at hand is exactly \lstinline|default|.
We can thus derive \lstinline|◇φ| as \lstinline|d⟨default⟩φ| and \lstinline|□φ| as \lstinline|d[default]φ|:
\begin{lstlisting}
instance [Unique β] [HasDynamicDiamond α β] : HasDiamond α where
  diamond φ := d⟨default⟩φ
instance [Unique β] [HasDynamicBox α β] : HasBox α where
  box φ := d[default]φ
\end{lstlisting}

This hierarchy makes the expected specialised notation available automatically from the generic modal operators together with the relevant structural properties of the signature.
On top of bare convenience, this makes notation uniform across logics, thus enabling reuse of definitions and theorems. We immediately leverage this in the next sections.

\subsection{Unary Modal Signatures}\label{sec:unary-signatures}

Say that a modal language has a signature where all modal operators are unary, i.e., they take exactly one argument.
We formalise this property as follows.
\begin{lstlisting}
(*@\lstlabel{PFunctor.Unary}@*)class PFunctor.Unary (P : PFunctor) where
  unary (a : P.A) : Unique (P.B a)
\end{lstlisting}

Under the assumption that a signature \lstinline|τ| is unary, we can develop specialised APIs and theorems that hide the full complexity of our general framework to end users. We give some illustrative examples.

Since we know that modal operators take exactly one argument, we can define simplified syntactic interfaces that do not expose proposition maps. For example, the case for applying a modal operator in a propositional context can just have a hole:
\begin{lstlisting}[end={,}]
@[match_pattern]
def Proposition.Context.dynDiamond (c : Context τ Atom) (a : τ.A) :
    Context τ Atom :=
  .triangle a default c fun ⟨i, hi⟩ => (hi (Subsingleton.elim i default)).elim
\end{lstlisting}
where the \lstinline|match_pattern| attribute registers this as a pattern for pattern matching, providing users with a way to deal directly with this specific case.

More interestingly, we can prove semantic characterisations that exhibit the behaviours of our derived operators:
\begin{lstlisting}[end={.}]
(*@\lstlabel{Satisfies.dynDiamond_iff_exists}@*)theorem Satisfies.dynDiamond_iff_exists {m : Model World τ Atom}
    {φ : Proposition τ Atom} : ⇓Modal[m,w ⊨ d⟨a⟩φ] ↔
      ∃ w', m.toFrame.diagonal a w w' ∧ ⇓Modal[m,w' ⊨ φ]
(*@\lstlabel{Satisfies.dynBox_iff_forall}@*)theorem Satisfies.dynBox_iff_forall {m : Model World τ Atom}
    {φ : Proposition τ Atom} : ⇓Modal[m,w ⊨ d[a]φ] ↔
      ∀ w', m.toFrame.diagonal a w w' → ⇓Modal[m,w' ⊨ φ]
\end{lstlisting}
Both of these theorems are proven by a single call to \lstinline|grind|, which is on purpose: initial failures at deriving these theorems automatically provided precious information on how we should extend and formulate some of our general results.

We provide a canonical constructor for a unary signature from a type of modal
operators, which uses \lstinline|Unit| as the type of argument positions:
\begin{lstlisting}[end={.}]
def PFunctor.mkUnary (A : Type u) : PFunctor where
  A := A
  B := fun _ => Unit
\end{lstlisting}
For any type of operators \lstinline|A|, \lstinline|PFunctor.mkUnary A| is an instance of \lstref{PFunctor.Unary} by construction.

We also introduce an associated constructor for frames with unary signatures, which we will apply in \Cref{sec:btl,sec:hml}:
\begin{lstlisting}[end={.}]
def ofRelations {τ : PFunctor} [τ.Unary] (r : τ.A → World → World → Prop) :
    Frame World τ where
  r i w ws := r i w (ws default)
\end{lstlisting}

\subsection{Basic Modal Signatures}\label{sec:basic-operators}
Extending the idea from the previous section, assume that a modal language has a signature \lstinline|τ| that is unary (\lstinline|Unary τ|) and has exactly one operator (\lstinline|Unique τ.A|).
This is exactly the signature of the basic modal language. Indeed, the first assumption gives us access to the results in \Cref{sec:unary-signatures} and the second makes available the notation for \lstinline|◇| and \lstinline|□| from \Cref{sec:derived-notation}.
In the rest of this section, statements implicitly assume \lstinline|Unary τ| and \lstinline|Unique τ.A|.

In such a \emph{unimodal} setting we can have a much simpler view on frames, since the diagonal of a frame for the default modal operator is always a binary relation:
\begin{lstlisting}[end={.}]
def rel (f : Frame World τ) : World → World → Prop := f.diagonal default
\end{lstlisting}
We use \lstinline|Frame.rel| to formalise a series of characterisation results, including the semantic characterisations of \lstinline|◇| and \lstinline|□|:
\begin{lstlisting}[end={,}]
theorem Satisfies.diamond_iff_exists {m : Model World τ Atom}
    {φ : Proposition τ Atom} :
    ⇓Modal[m,w ⊨ ◇φ] ↔ ∃ w', m.rel w w' ∧ ⇓Modal[m,w' ⊨ φ]
theorem Satisfies.box_iff_forall {m : Model World τ Atom}
    {φ : Proposition τ Atom} :
    ⇓Modal[m,w ⊨ □φ] ↔ ∀ w', m.rel w w' → ⇓Modal[m,w' ⊨ φ]
\end{lstlisting}
which are exactly as expected.
It is noteworthy that the proofs of these two results are exactly, respectively, the previously established theorems \lstref{Satisfies.dynDiamond_iff_exists} and \lstref{Satisfies.dynBox_iff_forall} from \Cref{sec:unary-signatures}. No new semantic argument is required: the unimodal interface inherits the semantics of the unary one.

It is convenient to have a universe-polymorphic constructor that builds a unimodal signature from an accessibility relation, which we will use in \Cref{sec:applied-modal-reasoning}.
\begin{lstlisting}
def τUnimodal (_ : Type u) : PFunctor.{u,u} where
  A := PUnit
  B := fun _ => PUnit
\end{lstlisting}

\subsection{Modal Logic for Lean}\label{sec:modal-logic-for-lean}
So far, we have used Lean to formalise modal logic.
Here, we change direction and investigate how modal logic can be used as a reasoning tool for ordinary Lean statements.

Given a modal frame over a Lean type \lstinline|α|, we can construct a modal model whose atomic propositions are simply predicates on \lstinline|α|:
\begin{lstlisting}[end={.}]
abbrev Model.ofPredicates (f : Frame α τ) : Model α τ (α → Prop) where
  toFrame := f
  v w P := P w
\end{lstlisting}
Thus, a Lean predicate \lstinline|P : α → Prop| can be used directly as an atomic modal proposition, and its modal semantics agrees definitionally with ordinary predicate application:
\begin{lstlisting}[end={.}]
(*@\lstlabel{Satisfies.ofPredicates_atom_iff}@*)theorem Satisfies.ofPredicates_atom_iff {P : α → Prop} (f : Frame α τ) :
    ⇓Modal[ofPredicates f,a ⊨ P] ↔ P a
\end{lstlisting}
This gives us a direct interface between modal reasoning and ordinary Lean propositions. The worlds of the modal model are Lean objects, while atomic propositions are properties of those objects.

The same idea applies to container-like structures. If \lstinline|β| carries a membership relation for elements of \lstinline|α|, we can use elements of \lstinline|β| themselves as modal atoms:
\begin{lstlisting}[end={.}]
abbrev Model.ofContainers [Membership α β] (f : Frame α τ) :
    Model α τ β where
  toFrame := f
  v w p := w ∈ p
\end{lstlisting}
Sets are the canonical instance of this construction.

These models let modal propositions characterise interactions of relations and predicates. For example, preservation of a predicate map by a frame is equivalent to the corresponding modal implication holding at every world:
\begin{lstlisting}
def Frame.PreservesMap (f : Frame α τ) (op : τ.A) (P : α → Prop)
    (Ps : τ.B op → α → Prop) : Prop :=
  ∀ w ws, f.r op w ws → P w → ∃ i, Ps i (ws i)
(*@\lstlabel{Satisfies.ofPredicates_preservesMap_iff}@*)theorem Satisfies.ofPredicates_preservesMap_iff {P : α → Prop} 
    {Ps : τ.B op → α → Prop} (f : Frame α τ) :
    (∀ w, ⇓Modal[ofPredicates f,w ⊨
      P → ∇[op](fun i => (Ps i : Proposition τ (α → Prop)))]) ↔
      f.PreservesMap op P Ps
\end{lstlisting}

These interfaces also let us reuse the logical-equivalence infrastructure from \Cref{sec:logical-equivalence}. In the predicate model induced by a frame, modal logical equivalence coincides with equality of denotations:
\begin{lstlisting}[end={,}]
(*@\lstlabel{Proposition.equivOfPredicates_denotation_eq}@*)theorem
    Proposition.equivOfPredicates_denotation_eq {f : Frame α τ}
    {φ₁ φ₂ : Proposition τ (α → Prop)} :
    (φ₁ ≡[Equiv.OfPredicates f] φ₂) ↔
      φ₁.denotation (ofPredicates f) = φ₂.denotation (ofPredicates f)
\end{lstlisting}
where \lstinline|Equiv.OfPredicates| is a shorthand for \lstinline|Proposition.Equiv (ofPredicates f)|.
This result is inherited directly from the generic characterisation of logical equivalence by denotational equality established in
\Cref{sec:logical-equivalence}.

More concretely, ordinary operations on Lean predicates can be related to their modal counterparts. For example, pointwise conjunction of two predicates is logically equivalent to modal conjunction:
\begin{lstlisting}[end={.}]
(*@\lstlabel{Proposition.equivOfPredicates_and}@*)theorem
    Proposition.equivOfPredicates_and
    {f : Frame α τ} {P Q : α → Prop} :
    (fun a => P a ∧ Q a) ≡[Equiv.OfPredicates f] (P ∧ Q)
\end{lstlisting}
Thus, when reasoning in the predicate model, we can move between ordinary Lean propositions and corresponding modal propositions without reproving their semantics.

The container interface admits analogous results. For example, whenever membership in the infimum of two containers agrees with conjunction of membership, their infimum is logically equivalent to modal conjunction:
\begin{lstlisting}[end={.}]
(*@\lstlabel{Proposition.ofContainers_inf_equiv}@*)theorem
    Proposition.ofContainers_inf_equiv
    [Membership α β] [Min β] (f : Frame α τ) (p q : β)
    (h : ∀ x, x ∈ p ⊓ q ↔ x ∈ p ∧ x ∈ q) :
    (↑(p ⊓ q) : Proposition τ β) ≡[Equiv.OfContainers f] (p ∧ q)
\end{lstlisting}

These constructions provide thus a general bridge between modal reasoning and existing Lean developments, which we will apply in \Cref{sec:applied-modal-reasoning}.

\section{Deriving Modal Logics in \cslib}
\label{sec:deriving-modal-logics}

In this section we put our framework to work, showing how different modal logics can be derived from our generic development by specialising operator signatures and reusing the interfaces, metatheory, and automation from the previous sections.

We consider three examples, which cover progressively richer cases.
\begin{itemize}
\item Basic modal logic arises from a unimodal signature, with a single unary
operator and hence a single accessibility relation.
\item Basic temporal logic uses two unary operators, corresponding to different temporal
relations.
\item Hennessy--Milner Logic ($\HML$) goes further by allowing an arbitrary family of unary operators indexed by transition labels.
\end{itemize}

The HML development also illustrates how our framework can be integrated with semantic structures from computer science. In
particular, we leverage that a Labelled Transition System (LTS) can be represented as a modal frame, so that our entire framework becomes directly available for reasoning about transition systems.

Together, these examples illustrate how the same infrastructure supports
unimodal, finitely multimodal, and label-indexed multimodal reasoning, as well as
its application to another existing semantic model, without introducing separate developments for each case.

\subsection{Basic Modal Logic}\label{sec:basic-modal-logic}

Basic modal logic is the logic that arises from the basic, unimodal signature (\Cref{sec:basic-operators}).
We use `the' signature informally: there are
many concrete choices of a unary signature with exactly one operator, but they
are all isomorphic.

Having a unimodal signature means automatically inheriting the basic modal operators and their semantics from \Cref{sec:basic-operators}.
We use our infrastructure to prove the well-known axioms from the literature and their accompanying characterisation results on frame conditions, given next. Frame conditions are formalised using our \lstinline|Frame.rel|; for example, we informally say the frame \lstinline|f| is reflexive for \lstinline|Std.Refl f.rel|.
\begin{description}
\item[$\axK$:] \lstinline|□(φ₁ → φ₂) → (□φ₁ → □φ₂)|.
\item[$\axDual$:] \lstinline|◇φ ↔ ¬□¬φ|.
\item[$\axT$:] \lstinline|φ → ◇φ| iff the frame is reflexive.
\item[$\axB$:] \lstinline|φ → □◇φ| iff the frame is symmetric.
\item[$4$:] \lstinline|◇◇φ → ◇φ| iff the frame is transitive.
\item[$5$:] \lstinline|◇φ → □◇φ| iff the frame is right Euclidean.
\item[$\axD$:] \lstinline|□φ → ◇φ| iff the frame is serial.
\item[$.2$:] \lstinline|◇□φ → □◇φ| iff the frame has the diamond property.
\end{description}
The proofs of all axioms under their frame conditions are straightforward thanks to our \lstinline|modal| grind set. The proofs of the only if directions require constructing a suitable valuation that exposes a violation of the corresponding frame condition, after which the remaining reasoning can again be handled by
\lstinline|grind|. These constructions follow the standard arguments from the
literature \cite{BRV01}.

We also prove further interaction laws for the modal operators. For example, formula \lstinline|◇(φ₁ ∧ φ₂) → (◇φ₁ ∧ ◇φ₂)| is valid on all frames.

\subsubsection{The Modal Cube}

Our infrastructure also lets us recover the \emph{modal cube}, which organises fifteen specialisations of basic modal logic
according to conditions imposed on the accessibility relation.

We define each logic semantically as the set of propositions valid in the
corresponding class of models. For example,
\begin{lstlisting}[end={.}]
def K World Atom := logic (Set.univ (α := Model World Atom))
def T World Atom := logic {m : Model World Atom | Std.Refl m.rel}
def S4 World Atom :=
  logic {m : Model World Atom | Std.Refl m.rel ∧ IsTrans World m.rel}
def S5 World Atom :=
  logic {m : Model World Atom | Std.Refl m.rel ∧ IsTrans World m.rel ∧ Relation.RightEuclidean m.rel}
\end{lstlisting}
The remaining nodes of the cube are defined analogously by combining
seriality, symmetry, transitivity, and right Euclideanness. Altogether we formalise
the fifteen logics of the cube.

The ordering of the cube follows directly from the semantic definition of \lstinline|logic| developed in \Cref{sec:axioms}. Recall that
\lstinline|logic| is antitone in its model class. Consequently, most inclusions between nodes of the cube amount to forgetting frame conditions, which is straightforward.
For example,
\begin{lstlisting}[end={.}]
theorem k_subset_t : (K World Atom ⊆ T World Atom)
\end{lstlisting}
Other inclusions additionally exploit standard implications between frame
conditions. For instance, every reflexive relation is serial, and therefore
the class of reflexive models is contained in the class of serial models.

Finally, the correspondence results we have established for axioms in basic modal logic immediately supply the characteristic validities at the appropriate nodes through simple \lstinline|grind| calls. For example,
\begin{lstlisting}[end={.}]
theorem K.k_valid :
    (□(φ₁ → φ₂) → (□φ₁ → □φ₂) : Proposition Atom) ∈ K World Atom
theorem T.t_valid : (φ → ◇φ : Proposition Atom) ∈ T World Atom
\end{lstlisting}

Thus, the modal cube is not formalised as a separate collection of bespoke modal systems. Its nodes arise uniformly by considering appropriate model classes over the same unimodal language, their ordering follows from generic properties, and the placement of the expected characteristic axioms comes from the correspondence and automation infrastructure developed previously.

\subsection{Basic Temporal Logic}\label{sec:btl}

Basic Temporal Logic ($\BTL$), also known as tense logic $\mathsf{K}_t$, has two unary modal operators, corresponding to operators expressing properties of future and past states. 
A basic presentation of the logic can be found in \cite{BRV01}; more details can be found in \cite{goranko2024temporal,gabbay1980temporal}.
We represent the future and past directions by the type
\begin{lstlisting}[end={,}]
inductive Direction where
  | future
  | past
\end{lstlisting}
and obtain the corresponding operator signature using the canonical
construction for unary signatures introduced in \Cref{sec:unary-signatures}:
\begin{lstlisting}[end={.}]
abbrev τBTL : PFunctor := PFunctor.mkUnary Direction
\end{lstlisting}
$\BTL$ propositions are therefore simply propositions of our generic modal
language over this signature:
\begin{lstlisting}[end={.}]
abbrev BTL.Proposition Atom := Modal.Proposition τBTL Atom
\end{lstlisting}

We expose the conventional temporal operators as specialised notation for the
generic unary modal operators:
\begin{lstlisting}
scoped prefix:40 "F" => fun φ => d⟨Direction.future⟩φ
scoped prefix:40 "P" => fun φ => d⟨Direction.past⟩φ
scoped prefix:40 "G" => fun φ => d[Direction.future]φ
scoped prefix:40 "H" => fun φ => d[Direction.past]φ
\end{lstlisting}
As usual, \lstinline|F| and \lstinline|P| read `sometime in the future' and
`sometime in the past', respectively, while their duals \lstinline|G| and
\lstinline|H| mean `always in the future' and `always in the past' (mnemonics: `it is always Going to be the case' and `it Has always been the case'). 

For the semantics of $\BTL$, it is important to insist that the accessibility relations for future and past are inverses of each other \cite{BRV01}.
We introduce a simple API that guarantees this by construction.
First, we define a $\BTL$ model as a relation \lstinline|r| together with a valuation:
\begin{lstlisting}[end={.}]
structure BTL.Model World Atom where
  r : World → World → Prop
  v : World → Atom → Prop
\end{lstlisting}
We then define a translation from $\BTL$ models into models of our generic framework,
\begin{lstlisting}[end={,}]
def BTL.Model.toModal (m : Model World Atom) : Modal.Model World τBTL Atom := ⟨m.toFrame, m.v⟩
\end{lstlisting}
where \lstinline|m.toFrame| interprets the future operator with \lstinline|r| and the past operator with its converse:
\begin{lstlisting}[end={.}]
def BTL.Model.toFrame (m : Model World Atom) : Frame World τBTL :=
  Frame.ofRelations fun
    | .future => m.r
    | .past => flip m.r
\end{lstlisting}
Thus, no new syntax or satisfaction relation is required.
In fact, we can define a notation for $\BTL$ semantic judgements,
\begin{lstlisting}[end={,}]
scoped notation "BTL[" m "," w " ⊨ " φ "]" =>
  Modal[BTL.Model.toModal m,w ⊨ φ]
\end{lstlisting}
so that we can write judgements like \lstinline|BTL[m,w ⊨ φ]| where \lstinline|m| is a $\BTL$ model.

As with basic modal logic, the expected semantic characterisations of the $\BTL$ operators are inherited directly from the generic unary development.
\begin{lstlisting}
theorem Satisfies.btl_future_iff_exists :
    ⇓BTL[m,w ⊨ F φ] ↔ ∃ w', m.r w w' ∧ ⇓BTL[m,w' ⊨ φ]
theorem Satisfies.btl_past_iff_exists :
    ⇓BTL[m,w ⊨ P φ] ↔ ∃ w', m.r w' w ∧ ⇓BTL[m,w' ⊨ φ]
theorem Satisfies.btl_always_iff_forall :
    ⇓BTL[m,w ⊨ G φ] ↔ ∀ w', m.r w w' → ⇓BTL[m,w' ⊨ φ]
theorem Satisfies.btl_hasAlwaysBeen_iff_forall :
    ⇓BTL[m,w ⊨ H φ] ↔ ∀ w', m.r w' w → ⇓BTL[m,w' ⊨ φ]
\end{lstlisting}
These results are proved directly by the corresponding generic theorems \lstref{Satisfies.dynDiamond_iff_exists} and
\lstref{Satisfies.dynBox_iff_forall} from
\Cref{sec:unary-signatures}.

The fact that the future and past relations are inverses also yields interaction laws between the temporal operators. For example:
\begin{lstlisting}[end={.}]
theorem Satisfies.btl_future_past : ⇓BTL[m,w ⊨ φ → G P φ]
theorem Satisfies.btl_past_future : ⇓BTL[m,w ⊨ φ → H F φ]
\end{lstlisting}
The first states that, if \lstinline|φ| holds now, then at every future world it
will be the case that \lstinline|φ| held sometime in the past; the second is its
past--future counterpart. Both are inherited from generic results in our general framework about pairs of modalities whose (diagonals of their) accessibility relations are inverse.

$\BTL$ therefore illustrates a different form of reuse from basic modal logic:
rather than collapsing to a single accessibility relation, it retains a small
multimodal signature whose operators satisfy a structural relationship at the
frame level. The generic unary syntax, semantics, metatheory, and automation
remain unchanged.

\subsection{Hennessy--Milner Logic}\label{sec:hml}

Our final example is Hennessy--Milner Logic ($\HML$), from \cite{hennessy1980observing}, a well-known logic for reasoning
about the behaviour of labelled transition systems.

Labelled transition systems (LTS) are already part of \cslib's general semantic infrastructure, independently of modal logic. They are defined as follows:
\begin{lstlisting}[end={.}]
structure LTS (State : Type u) (Label : Type v) where
  Tr : State → Label → State → Prop
\end{lstlisting}
Thus, an \lstinline|LTS State Label| consists of a transition relation
\lstinline|Tr|, where \lstinline|Tr s μ s'| states that \lstinline|s| can
perform a transition labelled \lstinline|μ| to \lstinline|s'|.

\cslib already included a formalisation of $\HML$ before the development
presented in this paper \cite{MPR26}.
That formalisation treated $\HML$ directly
over labelled transition systems and did not include propositional atoms. For each transition label \lstinline|μ|, one can write \lstinline|d⟨μ⟩φ| for `there exists a transition with label \lstinline|μ| from the current state to some state that satisfies \lstinline|φ|' and \lstinline|d[μ]φ| for `any state that the current state can reach through a transition with label \lstinline|μ| satisfies \lstinline|φ|'.
The formalisation in \cite{MPR26} culminates with a proof of the Hennessy--Milner theorem: under any image-finite LTS, two states are theory equivalent if and only if they are bisimilar.\footnote{An LTS is image-finite if, for any state \texttt{s} and transition label \texttt{μ}, \texttt{s} has only finitely many outgoing transitions with label \texttt{μ}. Two states are bisimilar if they can follow each other's transitions and always end up in bisimilar states (the definition is coinductive).}
As such, it represents a significant test of the capabilities and generality of our development.

To reconstruct $\HML$ in our framework, we define a translation from LTS into our frames that views every transition label as a binary accessibility relation:
\begin{lstlisting}[end={.}]
def LTS.toFrame : Frame State (mkUnary Label) :=
  Frame.ofRelations (fun μ s s' => lts.Tr s μ s')
\end{lstlisting}
Taking transition labels themselves as modal operators, the $\HML$ signature is
\lstinline|mkUnary Label|, and $\HML$ propositions are simply:
\begin{lstlisting}[end={,}]
abbrev HML.Proposition Label Atom := Modal.Proposition (mkUnary Label) Atom
\end{lstlisting}
which immediately gives us access to the notations \lstinline|d⟨μ⟩φ| and \lstinline|d[μ]φ|.
Note that this also extends the previous development from \cite{MPR26} to having atomic propositions and valuations.

Proceeding to semantics, we want to keep exposing users of HML to LTS as usual rather than our general frames. We follow the same idea as for the basic temporal language. That is, we define a dedicated structure,
\begin{lstlisting}[end={,}]
structure HML.Model State Label Atom where
  lts : LTS State Label
  v : State → Atom → Prop
\end{lstlisting}
its translation into models of our generic framework,
\begin{lstlisting}[end={.}]
def HML.Model.toModal (m : Model State Label Atom) :=
  Modal.Model.mk m.lts.toFrame m.v
\end{lstlisting}
and accompanying notation for semantic judgements in $\HML$,
\begin{lstlisting}[end={.}]
scoped notation "HML[" m "," s " ⊨ " φ "]" =>
  Modal[HML.Model.toModal m,s ⊨ φ]
\end{lstlisting}

We can now prove the expected semantics of $\HML$ modal operators as characterisation results:
\begin{lstlisting}[end={.}]
theorem Satisfies.hml_dynDiamond_iff_exists :
    ⇓HML[m,s ⊨ d⟨μ⟩φ] ↔ ∃ s', m.lts.Tr s μ s' ∧ ⇓HML[m,s' ⊨ φ]
theorem Satisfies.hml_dynBox_iff_forall :
    ⇓HML[m,s ⊨ d[μ]φ] ↔ ∀ s', m.lts.Tr s μ s' → ⇓HML[m,s' ⊨ φ]
\end{lstlisting}

These semantic characterisations are sufficient to recover the main
metatheory of the previous $\HML$ development with only minor refactoring. In particular, we obtain the Hennessy--Milner theorem: for any image-finite LTS, if valuation is preserved by bisimilarity, then theory equivalence coincides with bisimilarity.\footnote{The \texttt{HomBisimilarity} relation mentioned in the theorem statement stands for `homogeneous bisimilarity'; that is, bisimilarity for states in the same LTS.}
\begin{lstlisting}
theorem theoryEq_eq_bisimilarity
    [image_finite : ∀ s μ, Finite (m.lts.image s μ)]
    (hv : ∀ {s1 s2}, s1 $\sim$[m.lts] s2 → ∀ p, m.v s1 p ↔ m.v s2 p) :
  TheoryEq m.toModal = HomBisimilarity m.lts
\end{lstlisting}
Our condition \lstinline|hv| is new, because our version of $\HML$ includes propositional atoms whereas the previous \cslib development does not. In the atom-free fragment corresponding to the previous development, this condition is vacuous, so the original Hennessy–Milner theorem is recovered.

\section{Applied Modal Reasoning: Mathematics and Computer Science}\label{sec:applied-modal-reasoning}

In this section, we illustrate some of the opportunities that arise from having a general, extensive, and reusable formalisation of modal logic available inside Lean.

We consider three case studies from mathematics and computer science. They are not intended to be exhaustive; rather, they demonstrate how our framework can be connected to existing formal developments and used to establish results about concrete mathematical and computational structures.

\subsection{Mathematics: Scalar Multiplication and Radicals of Ideals}\label{sec:smul}

Our first case study illustrates how modal reasoning can be applied directly to ordinary mathematical structures, thanks to the modal interface for Lean predicates and containers built in \Cref{sec:modal-logic-for-lean}.
We begin by studying a general construction for scalar multiplication operators, and then instantiate it to positive exponentiation to reason about radicals of ideals.\footnote{In Mathlib, an ideal is an additive submonoid closed under multiplication. The radical of an ideal \lstinline|I| is the ideal consisting of exactly the elements \lstinline|r| such that \lstinline|r^n ∈ I| for some natural number \lstinline|n|.}

Any scalar multiplication induces a binary relation: an element \lstinline|y| is accessible from \lstinline|x| whenever it can be obtained by multiplying \lstinline|x| by some scalar. Formally,
\begin{lstlisting}[end={.}]
def Relation.ofSMul (M α : Type*) [SMul M α] (x y : α) : Prop :=
  ∃ m : M, m • x = y
\end{lstlisting}
In Mathlib, the class \lstinline|SMul| for scalar multiplication is extended to define actions with certain properties.
We prove that algebraic properties of the action induce familiar frame properties.
\begin{itemize}
\item Monoid actions give reflexive relations.
\item Group actions give symmetric relations.
\item Semigroup actions give transitive relations.
\item Actions with a nonempty type of scalars give serial relations.
\item Actions that commute with themselves give relations with the diamond property.
\end{itemize}
By this development, existing algebraic structures and their properties can immediately be exposed to our modal framework.
In particular, we can take the binary relation induced by a scalar multiplication operator in this way as the accessibility relation in basic modal logic.

Let now \lstinline|R| be a commutative semiring and take exponentiation by positive natural numbers as the action; that is, we instantiate \lstinline|n • x| as \lstinline|x^n|.
This action induces the accessibility relation \lstinline|PosPow R|:
\begin{lstlisting}[end={.}]
abbrev PosPow α [Monoid α] := Relation.ofSMul ℕ+ α
\end{lstlisting}
We construct a container model where ideals \emph{are} the atoms:
\begin{lstlisting}[end={.}]
abbrev idealPowerFrame [Semiring R] : Frame R (τUnimodal R) :=
  Frame.ofRelation (PosPow R)
abbrev idealPowerModel {R : Type u} [Semiring R] :
    Model R (τUnimodal R) (Ideal R) := Model.ofContainers idealPowerFrame
\end{lstlisting}
In this model, the worlds are elements of the semiring, atomic propositions are ideals, and satisfaction of an ideal atom is ordinary ideal membership.
Moreover, since ideals are closed under positive powers, ideal membership is preserved by the accessibility relation.

The key observation is that the radical of an ideal is precisely the denotation of its modal possibility:
\begin{lstlisting}[end={,}]
(*@\lstlabel{Ideal.radical_eq_modal_denotation}@*)theorem Ideal.radical_eq_modal_denotation [CommSemiring R] (I : Ideal R) :
    (I.radical : Set R) = Proposition.denotation idealPowerModel (◇I)
\end{lstlisting}
because membership in \lstinline|◇I| means that some positive power of the current world belongs to \lstinline|I|. Equivalently, in this model the radical of an ideal is logically equivalent to modal possibility:
\begin{lstlisting}[end={,}]
(*@\lstlabel{Ideal.radical_equiv_diamond}@*)theorem Ideal.radical_equiv_diamond [CommSemiring R] (I : Ideal R) :
    (I.radical : Proposition (τUnimodal R) (Ideal R)) ≡[IdealEquiv] ◇I
\end{lstlisting}
where \lstinline|IdealEquiv| is a shorthand for \lstinline|Proposition.Equiv idealPowerModel|.
Because of this correspondence, we can apply modal reasoning to obtain algebraic laws about radicals. We demonstrate this by reproving two results in Mathlib using modal logic.

\subsubsection{Idempotence via Axioms 4 and T}\label{sec:radical-idem}
First, idempotence of the radical operation follows from standard modal laws enabled by the fact that the frame induced by positive exponentiation is reflexive and transitive:
\begin{lstlisting}[end={.}]
(*@\lstlabel{Ideal.radical_idem}@*)theorem Ideal.radical_idem [CommSemiring R] (I : Ideal R) :
    I.radical.radical = I.radical
\end{lstlisting}
In the proof, we apply \lstref{Ideal.radical_eq_modal_denotation} and \lstref{Proposition.denotation_eq_of_equiv} to switch to modal reasoning; this reduces the goal to
\begin{lstlisting}[end={.}]
 ◇(I.radical) ≡[IdealEquiv] ◇I
\end{lstlisting}
Henceforth, the proof is purely modal.
Since logical equivalence is a congruence, we can rewrite the left-hand side by \lstref{Ideal.radical_equiv_diamond} to obtain the goal
\begin{lstlisting}[end={,}]
 ◇◇I ≡[IdealEquiv] ◇I
\end{lstlisting}
which is exactly the modal equivalence induced by Axioms 4 and T
(\Cref{sec:basic-modal-logic}): transitivity gives
$\Diamond\Diamond I \to \Diamond I$, while reflexivity gives
$\Diamond I \to \Diamond\Diamond I$.

\subsubsection{Distribution over Intersection via Intersection as Conjunction}\label{sec:radical-dist}

The second result concerns intersection. For ideals, the modal proposition expressing reachability into an intersection is equivalent to the conjunction of the two corresponding possibilities:
\begin{lstlisting}[end={.}]
(*@\lstlabel{Ideal.inf_modelEquiv}@*)theorem Ideal.inf_modelEquiv {R : Type u} [Semiring R] (I J : Ideal R) :
    (◇(I ⊓ J : Ideal R) : Proposition (τUnimodal R) (Ideal R))
      ≡[IdealEquiv] ( ◇I ∧ ◇J )
\end{lstlisting}
This is an instance of a more general modal result that we prove separately, which states that \lstinline|◇(φ₁ ∧ φ₂) ≡[Equiv m] ◇φ₁ ∧ ◇φ₂| whenever the accessibility relation is transitive, has the diamond property, and preserves both \lstinline|φ₁| and \lstinline|φ₂|.
For positive exponentiation, these hypotheses follow from the general results on scalar multiplication established at the beginning of \Cref{sec:smul}: the action is transitive, commutes with itself, and preserves membership in ideals.

Combining this result with the characterisation of radicals by modal denotations, we obtain distribution of radical over intersection:
\begin{lstlisting}[end={.}]
(*@\lstlabel{Ideal.radical_inf}@*)theorem Ideal.radical_inf [CommSemiring R] (I J : Ideal R) :
    (I ⊓ J).radical = I.radical ⊓ J.radical
\end{lstlisting}

Let us take stock after \Cref{sec:radical-idem,sec:radical-dist}. The significance of these examples is not, of course, that radicals require modal logic for their formalisation. Rather, they show that, once ordinary mathematical operations are exposed through suitable accessibility relations, our modal infrastructure can become a proof tool for mathematics that was not developed with modal logic in mind. Here, properties of scalar actions, generic modal laws, logical equivalence, congruence, and denotational reasoning have been combined to recover ordinary theorems about radicals of ideals.

\subsection{Theory of Programming Languages: $\lambda$-calculus and Type Safety}
\label{sec:type-safety}

Our second case study applies modal reasoning to metatheory of programming languages.
We use the simply typed $\lambda$-calculus (STLC), which is already formalised independently in \cslib \cite{cslib-spine}.
The remainder of this section assumes some familiarity with the simply typed $\lambda$-calculus and its terminology.

The case study deals with \emph{type safety} for STLC, a well-known property stating that a well-typed term (representing a program) can only reach (possibly in multiple execution steps) either \emph{values} (terms that represent finished computations) or terms that can keep running. In other words, a well-typed term can never get improperly `stuck'.
The standard strategy to achieve this result goes under the informal formula of `safety = progress + preservation' \cite{P02}; that is, one proves two simpler results.
\begin{description}
\item[Progress:] A well-typed term is either a value or can take another step.
\item[Preservation:] If a well-typed term can take a step, the result is well-typed.
\end{description}
We show that this argument can be captured through modal reasoning.

We define two models to reason about $\lambda$-calculus, which take $\lambda$-calculus terms as worlds: one where the accessibility relation is single-step full $\beta$-reduction, and the other where the accessibility relation is multistep full $\beta$-reduction. Atomic propositions are ordinary predicates on terms. The models are obtained directly through the predicate interface of \Cref{sec:modal-logic-for-lean}:
\begin{lstlisting}[end={.}]
def singleStepModel :
    Model (Term Var) (τ := τUnimodal (Term Var)) (Term Var → Prop) :=
  Model.unimodalOfPredicates (· ⭢βᶠ ·)
def multiStepModel :
    Model (Term Var) (τUnimodal (Term Var)) (Term Var → Prop) :=
  Model.unimodalOfPredicates (· ↠βᶠ ·)
\end{lstlisting}
Terms have type \lstinline|Term Var|, because they are parametric on the type of atomic variables. The choice of \lstinline|Var| is immaterial to our development, so it is an implicit parameter.

We then define abbreviations for the type of propositions in this model, as well as the propositions that a term has a given type, is a value, or is just a term (which is always true).
\begin{lstlisting}
abbrev LProposition (Var : Type u) :=
  Proposition (τUnimodal (Term Var)) (LAtom Var)
abbrev HasType (τ : Ty Base) : LProposition Var :=
  (Typing (Var := Var) [] · τ)
abbrev IsValue : LProposition Var := Term.Value (Var := Var)
abbrev IsTerm : LProposition Var := fun _ : Term Var => True
\end{lstlisting}
Thus, \lstinline|◇IsTerm| simply states that a term has a successor.

With our definitions in place, we can express the progress and preservation properties as modal propositions:
\begin{lstlisting}[end={,}]
(*@\lstlabel{progress_singleStep_modal}@*)theorem progress_singleStep_modal (τ : Ty Base) (t : Term Var) :
    ⇓Modal[singleStepModel,t ⊨ HasType τ → IsValue ∨ ◇IsTerm]
(*@\lstlabel{preservation_singleStepModel_modal}@*)theorem preservation_singleStepModel_modal (τ : Ty Base) (t : Term Var) :
    ⇓Modal[singleStepModel,t ⊨ HasType τ → □HasType τ]
\end{lstlisting}
where \lstinline|Ty Base| is the type of STLC types.
Their proofs simply lift the existing STLC progress and preservation theorems into the modal model.

We observe that the combination of progress and preservation has a generic modal formulation, independent of the $\lambda$-calculus:
\begin{lstlisting}[end={.}]
(*@\lstlabel{safety_of_preservation_progress}@*)theorem safety_of_preservation_progress
    {φ₁ φ₂ φ₃ φ₄ : Proposition (τUnimodal α) Atom}
    (hpres : ⇓Modal[m,w ⊨ φ₁ → □φ₂])
    (hprog : ⇓Modal[m,w ⊨ φ₁ → φ₃ ∨ ◇φ₄]) :
    ⇓Modal[m,w ⊨ φ₁ → φ₃ ∨ ◇φ₂]
\end{lstlisting}
The argument is purely modal. Preservation says that every successor satisfies
\lstinline|φ₂|, while progress supplies either \lstinline|φ₃| immediately or
some successor. In the latter case, necessity and possibility combine to give
a successor satisfying \lstinline|φ₂|.

Instantiating this principle with typing, values, and terms yields what we call `single-step type safety':
\begin{lstlisting}[end={.}]
(*@\lstlabel{type_safety_singleStep_modal}@*)theorem type_safety_singleStep_modal (τ : Ty Base) (t : Term Var) :
    ⇓Modal[singleStepModel,t ⊨ HasType τ → IsValue ∨ ◇HasType τ]
\end{lstlisting}
From here, we can reach the standard type safety result about multistep reductions remaining abstract of $\lambda$-calculus.
Since multistep reduction is the reflexive and transitive closure of single-step reduction, we can reuse a generic principle for basic modal logic about relations that preserve predicates, which gives us:
\begin{lstlisting}[end={.}]
(*@\lstlabel{preservation_multiStep_modal}@*)theorem preservation_multiStep_modal (τ : Ty Base) (t : Term Var) :
    ⇓Modal[multiStepModel,t ⊨ HasType τ → □HasType τ] :=
  Satisfies.unimodalOfPredicates_preserves_reflTransGen
    (preservation_singleStepModel_modal τ ·) t
\end{lstlisting}
To state standard type safety of every term reached by a multistep execution, we need to combine the single-step and multistep models. Since we are working in
unimodal logic, we reify single-step safety as an atomic predicate on terms:\footnote{Alternatively, we could have instantiated our framework with two modalities, one for single- and the other for multistep reductions.}
\begin{lstlisting}[end={.}]
abbrev IsSingleStepSafe (τ : Ty Base) : LProposition Var :=
  fun t : Term Var => ⇓Modal[singleStepModel,t ⊨ IsValue ∨ ◇HasType τ]
\end{lstlisting}
The modality occurring inside \lstinline|IsSingleStepSafe| is therefore interpreted using single-step reduction, whereas an outer modality in \lstinline|multiStepModel| ranges over finite executions.

We can thus express type safety in our framework as follows.
\begin{lstlisting}
(*@\lstlabel{type_safety_modal}@*)theorem type_safety_modal (τ : Ty Base) (t : Term Var) :
    ⇓Modal[multiStepModel,t ⊨ HasType τ → □IsSingleStepSafe τ]
\end{lstlisting}
It states that, from a well-typed initial term, every term reachable by a finite execution is single-step safe.
The proof consists of generic modal reasoning. Multistep preservation gives \lstinline|HasType τ → □HasType τ|, while single-step safety at each reachable term yields
\lstinline|□(HasType τ → IsSingleStepSafe τ)|.
Axiom K then combines these two facts to obtain \lstinline|□IsSingleStepSafe τ|. No further STLC-specific reasoning is required.

Finally, unfolding the semantics of the two models recovers the conventional type safety statement:
\begin{lstlisting}[end={.}]
(*@\lstlabel{type_safety}@*)theorem type_safety
    {t t' : Term Var} {τ : Ty Base}
    (ht : [] ⊢ t ∶ τ) (hsteps : t ↠βᶠ t') :
    t'.Value ∨ ∃ t'', t' ⭢βᶠ t'' ∧ [] ⊢ t'' ∶ τ
\end{lstlisting}
Hence every term reachable from a well-typed term is either already a value, or can take another reduction step to a well-typed term.
Interestingly, this result was missing from \cslib before our development.

\subsection{Concurrency Theory: CCS and Milner's Vending Machines}
\label{sec:ccs-vending}

Our third case study connects the Hennessy--Milner logic development of \Cref{sec:hml} with an independently formalised language for concurrency theory: Milner's Calculus of Communicating Systems (CCS) \cite{Milner80}. The operational semantics of CCS in \cslib is already presented as an LTS, so it can be used directly to construct an $\HML$ model.

We consider Milner's standard vending machine example, where a vending machine accepts a coin to get some coffee or tea. There are two versions: a deterministic version, which accepts a coin and then offers a choice between tea and coffee; and a nondeterministic version, which makes the choice when the coin is accepted.
In \cslib, the two machines are represented using recursive CCS constants:
\begin{lstlisting}[end={,}]
def vendingDefs : Constant → Option (Process String Constant)
  | .vm => some <| `(CCS| Coin. ((Tea. const .vm) + (Coffee. const .vm)))
def vendingDefsND : Constant → Option (Process String Constant)
  | .vm => some <| `(CCS| (Coin. Tea. const .vm) + (Coin. Coffee. const .vm))
\end{lstlisting}
where \lstinline|+| is the choice operator.
Their corresponding labelled transition systems are obtained directly from the generic CCS semantics.
\begin{lstlisting}
abbrev ltsD  := CCS.lts (defs := vendingDefs)
abbrev ltsND := CCS.lts (defs := vendingDefsND)
\end{lstlisting}

The two processes differ in their branching structure, and \cslib includes a proof that calling the \lstinline|vm| constant in the two LTSs does not give bisimilar states:
\begin{lstlisting}[end={.}]
(*@\lstlabel{vm_ltsD_ltsND_not_bisim}@*)theorem vm_ltsD_ltsND_not_bisim : ¬(vm $\sim$[ltsD, ltsND] vm)
\end{lstlisting}
The reason is that, after the initial coin action, the deterministic machine reaches a state from which both tea and coffee are available; by contrast, each coin transition of the nondeterministic machine commits to one of the two drinks.

We can expose the same distinction logically. Since this example does not require atomic propositions, we construct $\HML$ models with atom type \lstinline|Empty|:
\begin{lstlisting}
def modelD : HML.Model (Process String Constant) (Act String) Empty :=
  ⟨ltsD, fun _ p => nomatch p⟩
def modelND : HML.Model (Process String Constant) (Act String) Empty :=
  ⟨ltsND, fun _ p => nomatch p⟩
\end{lstlisting}
No CCS-specific satisfaction relation is needed, because the  two LTSs produced by the CCS semantics are interpreted by the generic $\HML$ semantics developed earlier.

Consider now the $\HML$ proposition
\begin{lstlisting}[end={,}]
(*@\lstlabel{choiceAfterCoin}@*)def choiceAfterCoin : HML.Proposition (Act String) Empty :=
  d[Coin](d⟨Tea⟩⊤ ∧ d⟨Coffee⟩⊤)
\end{lstlisting}
which states that after every coin transition, both a tea transition and a coffee transition remain possible.
The deterministic machine satisfies this proposition, whereas the nondeterministic machine does not:
\begin{lstlisting}[end={.}]
(*@\lstlabel{vm_modelD_choiceAfterCoin}@*)theorem vm_modelD_choiceAfterCoin : ⇓HML[modelD,vm ⊨ choiceAfterCoin]
(*@\lstlabel{vm_modelND_not_choiceAfterCoin}@*)theorem vm_modelND_not_choiceAfterCoin : ¬⇓HML[modelND,vm ⊨ choiceAfterCoin]
\end{lstlisting}
For example, it has a coin transition to the process
\lstinline|Tea. const .vm|, from which no coffee transition is available.
The proofs use the generic $\HML$ characterisations of dynamic box and diamond,
\lstinline|hml_dynBox_iff_forall| and
\lstinline|hml_dynDiamond_iff_exists|, together with the CCS transition
semantics.

This concrete example illustrates the distinguishing power characterised abstractly by \lstinline|theoryEq_eq_bisimilarity| in \Cref{sec:hml}: $\HML$ propositions can observe the same branching structure that is central to bisimilarity.

This example also exhibits clearly the advantage of integrating with an ecosystem like \cslib. The operational semantics of CCS and the theory of labelled transition systems were developed before our work. However, because their APIs are integrated, any concrete CCS system can be subjected directly to modal analysis in our framework.

\section{Discussion}
\label{sec:discussion}

In this section we discuss some informal insights that we have gained through our development, as well as interesting directions for future work.

\subsection{Lessons for Modal Logic}

Mechanising modal logic at this level of generality within \cslib's ecosystem has taught us something about the theory itself.

A first lesson is that formalisation is particularly effective at exposing the exact assumptions on which results depend. Throughout the development, hypotheses that are often left implicit could be isolated and weakened independently: for instance, worlds need not be inhabited for most results, modal operators need not have finite arity, and assumptions such as nonemptiness or decidable equality appear only where they are actually required. The same effect is visible in the applications. For example, for relations induced by scalar multiplication, reflexivity, transitivity, symmetry, seriality, and related modal properties follow from distinct algebraic assumptions on the action. The theorem prover thus serves not only to certify proofs, but to reveal the dependency structure of the theory.

The general treatment of modal similarity types in \cite{BRV01} proved remarkably robust as a foundation for this development. Many of its abstractions transferred directly to Lean, while the formal setting encouraged us to push the framework even further, in particular by allowing modal operators of arbitrary, potentially infinite arity. More generally, constructions that are left implicit, relegated to exercises, or treated as folklore in textbook presentations must become explicit when they are to support a reusable library. Pursuing this systematically led us to formulations of frame conditions at the level of arbitrary operators and to the diagonal construction used throughout our correspondence results. Formalisation can therefore be generative: insisting on definitions and proofs that work uniformly can expose useful structure that can be overlooked in informal developments.

A broader lesson concerns the role of modal logic itself. Modal logic is, at heart, a language for reasoning about relational structures, and this perspective becomes especially concrete inside a proof assistant. The examples of \Cref{sec:applied-modal-reasoning} show that modal propositions can package recurring relational principles arising in algebra, operational semantics, and transition systems, and that generic modal theorems can then guide proofs in those domains. This suggests that modal logic can serve not only as an object of formalisation, but also as a reusable and effective reasoning layer for mathematics and computer science.

Finally, our treatment of logical equivalence benefitted substantially from abstractions more familiar from concurrency theory and programming languages. In particular, making syntactic contexts and congruence explicit provided a useful discipline for structuring equivalence and substitution principles. This viewpoint complements the usual semantic presentation of modal logic: once equivalence is intended to function as a library interface, its interaction with contexts and operators becomes part of the theory that must be represented directly. This helped determine the form of the resulting modal library.

\subsection{Lessons for Library-Grade Formalisation}\label{sec:discussion-library-grade}

Our development was guided by two complementary forms of reuse. We sought \emph{vertical reuse}, so that specialised logics could reconstruct their syntax, semantics, metatheory, and characteristic results from more general layers; and \emph{horizontal reuse}, so that modal logic could be applied directly to independently developed structures such as algebraic actions, reduction systems, and labelled transition systems. Reaching these aims required several design iterations and influenced the development of the library throughout, from the choice of generic signatures and semantic interfaces to the separation between generic, unary, and unimodal layers.

\begin{table}
\centering
\begin{filecontents*}{generated/modal-stats.csv}
metric,hml_old,hml_new,hml_reduction,bml_old,bml_new,bml_reduction
Lines of code,686,224,67.3,980,596,39.2
Theorems/lemmas,43,12,72.1,69,39,43.5
Definitions,14,1,92.9,28,19,32.1
Abbreviations,2,1,50.0,4,2,50.0
Inductive types,2,0,100.0,3,0,100.0
Structures,2,1,50.0,3,0,100.0
Instances,19,0,100.0,19,6,68.4
\end{filecontents*}
\pgfplotstabletypeset[
  col sep=comma,
  string type,
  columns={
    metric,
    hml_old,
    hml_new,
    hml_reduction,
    bml_old,
    bml_new,
    bml_reduction
  },
  columns/metric/.style={
    column name={\textbf{Metric}},
    column type=l
  },
  columns/hml_old/.style={
    column name={Before},
    column type=r
  },
  columns/hml_new/.style={
    column name={After},
    column type=r
  },
  columns/hml_reduction/.style={
    column name={Reduction (\%)},
    column type=r
  },
  columns/bml_old/.style={
    column name={Before},
    column type=r
  },
  columns/bml_new/.style={
    column name={After},
    column type=r
  },
  columns/bml_reduction/.style={
    column name={Reduction (\%)},
    column type=r
  },
  every head row/.style={
    before row={
      \toprule
      & \multicolumn{3}{c}{\textbf{Hennessy--Milner Logic}}
      & \multicolumn{3}{c}{\textbf{Basic Modal Logic}}\\
      \cmidrule(lr){2-4}
      \cmidrule(lr){5-7}
    },
    after row=\midrule
  },
  every last row/.style={
    after row=\bottomrule
  }
]{generated/modal-stats.csv}

\caption{Logic-specific implementations before and after deriving the logics from the generic modal framework.}
\label{table:modal-reuse}
\end{table}

To give an idea of the effect of our design, we can compare the logic-specific implementations of $\HML$ and basic modal logic before and after the introduction of our generic modal framework -- the two modal logics that were present in \cslib before we developed our framework. 
The basic modal logic implementation is from the same authors of this article and was developed as part of this work as a first step towards the present framework.
The implementation of $\HML$ is from \cite{MPR26}. 
The comparison is given in \Cref{table:modal-reuse}.
First of all, we can observe substantial reductions in lines of code even though all characteristic results about the two logics have been preserved. While this is a coarse metric, the result for $\HML$ is particularly striking because the new implementation is more expressive -- it adds the feature of atomic propositions.
Looking at the code, we can observe that a lot of common structure has moved to reusable layers, making many specialised definitions and theorems unnecessary. The strong reductions in the number of instances is due to careful design of our classes and the development of a hierarchy of generic instances.

Our first lesson is therefore methodological: \emph{seek reuse explicitly}.
If vertical and horizontal reuse are treated as design objectives from the beginning, they shape where definitions and theorems should live, which
interfaces should be exposed, and which constructions should be made generic. In our experience, this made both reconstructing specialised modal logics and applying modal reasoning outside logic itself substantially more direct.
Whenever a logic was difficult to reconstruct or apply, we reconsidered the design of our generic definitions and classes.

A second lesson concerns automation. Lean's \texttt{grind} mechanism was
particularly effective in a development of this scale. It reduced the amount
of glue reasoning needed when moving between abstraction layers, and it also
allowed the library to expose useful downstream reasoning support through the
dedicated \texttt{modal} database. In this role, automation is not merely a
convenience for developing the library: it becomes part of the interface
offered to its users. A well-designed collection of automation rules can make
results inherited from generic layers available in specialised developments
without repeatedly exposing their internal encodings.

These two lessons reinforce each other. Reuse determines the abstraction boundaries of the library; automation helps make those boundaries effective in practice. For library-grade formalisation, we therefore see both as design concerns rather than as consequences to be considered only after the mathematics has been formalised.
Summarising: reuse and automation need to be first-class citizens, and Lean's toolkit has served this purpose well.

\subsection{Limitations and Future Work}

The framework developed in this paper opens several directions for extension. The literature on and applicability of modal logics are extensive. Thus, rather than attempting an exhaustive list, we highlight three broad categories that seem particularly important.

Our present development is predominantly semantic: it provides syntax, relational semantics, equivalence, frame correspondence, derived logics, and reusable reasoning infrastructure, but does not yet offer a comparable architecture for axiom systems, sequent calculi or other proof systems.
The most natural next step would thus be to extend the library with proof systems
and their corresponding metatheory, in the spirit of \cite{OBKS24,OML26}.
Developing this direction would hopefully enable us to establish soundness and completeness of the calculi in a uniform way across modal logics, and it would be interesting to investigate how much of the usual proof-theoretic machinery can be inherited along the same abstraction hierarchy used here for semantics.

Another interesting direction is extending our generic modal language. Two particularly important features would be fixpoints and quantification. Fixpoint operators would bring the modal $\mu$-calculus within scope, while quantification would allow for abstraction.
This would be a test both for our architecture and the theory of generic modal languages proposed in \cite{BRV01}.

Finally, a third direction regards automation. The current \lstinline|modal| database for \lstinline|grind| already shows that a substantial amount of routine reasoning can be hidden behind a small public interface, but the present setup is intentionally conservative. In particular, generic modal semantics are not unfolded indiscriminately, since doing so can quickly enlarge the search space.
Better automation would strengthen the reasoning offered to downstream developments.

\section{Conclusion}
\label{sec:conclusion}

We have presented a library-grade formalisation of modal logic in Lean, designed around reuse rather than around any single modal system. Starting from a generic polyadic language and relational semantics, we developed reusable metatheory, interfaces, and automation, and used them to derive specialised modal logics including basic modal logic, basic temporal logic, and Hennessy--Milner Logic.

Our experience suggests that formalising a logic for a library requires a different perspective from formalising it in isolation.
The relevant question is not only which definitions and theorems should be mechanised, but which abstractions will allow future developments to inherit, specialise, combine, and apply them.
To this end, our results demonstrate the practical utility of the textbook generic theory of modal logic \cite{BRV01}.
We hope that the architecture developed here provides a basis for extending the range of modal logics available in \cslib, while making modal reasoning itself increasingly useful in formalised mathematics and computer science.
More broadly, we hope that our approach, and the lessons learned in developing it, can inform the development of other logics as reusable components of the library.

\backmatter

\bmhead{Acknowledgements}
Co-funded by the European Union (ERC, CHORDS, 101124225). Views and opinions expressed are however those of the authors only and do not necessarily reflect those of the European Union or the European Research Council. Neither the European Union nor the granting authority can be held responsible for them.

\bibliography{biblio}





\end{document}

%% file: macros.tex
\newcommand{\modsim}{\tau}
\newcommand{\modL}[1]{\mathsf{ML}(#1)}
\newcommand{\tri}{\Delta}
\newcommand{\trid}{\nabla}

\newcommand{\nat}{\mathbb{N}}
\newcommand{\DIA}{\Diamond}
\newcommand{\BOX}{\Box}
\newcommand{\DIAp}[1]{\langle #1\rangle}
\newcommand{\BOXp}[1]{[ #1]}

\newcommand{\Prop}{\mathsf{Prop}}

\newcommand{\OR}{\lor}
\newcommand{\IMP}{\rightarrow}

\newcommand{\NEG}{\lnot}

\newcommand{\rel}[2]{#1 R #2}

\newcommand{\fr}{\mathfrak{F}}
\newcommand{\cl}{\mathcal{C}}
\newcommand{\model}{\mathcal{M}}
\newcommand{\logic}{\mathsf{Log}}

\newcommand{\BTL}{\mathsf{BTL}}
\newcommand{\HML}{\mathsf{HML}}

\newcommand{\K}{\mathsf{K}}

\newcommand{\sfive}{\mathsf{S5}}
\newcommand{\sfour}{\mathsf{S4}}

\newcommand{\axK}{\mathsf{k}}
\newcommand{\axT}{\mathsf{t}}
\newcommand{\axD}{\mathsf{d}}
\newcommand{\axDual}{\mathsf{dual}}
\newcommand{\axB}{\mathsf{b}}